\documentclass{aa} 
\usepackage{graphicx}
\usepackage{txfonts}
\usepackage{longtable}

\input epsf.sty

\begin{document}

\title{The mass of the black hole in RE J1034+396}

\author{B. Czerny \inst{1,2,3}
        \and
      B. You\inst{2}
      \and
        A. Kurcz \inst{4}
       \and 
        J. \' Sredzi\' nska \inst{2}
        \and
     K. Hryniewicz \inst{2}
      \and
     M. Niko\l ajuk \inst{5}
     \and
     M. Krupa \inst{4}
     \and
     J.-M. Wang \inst{3}
     \and
     C. Hu \inst{3}
     \and
      P. T. \. Zycki \inst{2}
        }

 \offprints{B. Czerny}

\institute{Center for Theoretical Physics, Polish Academy of Sciences, Al. Lotnik\' ow 32/46, 02-668 Warsaw, Poland\\
     \and Copernicus Astronomical Center, Polish Academy of Science, Bartycka 18, 00-716 Warsaw, Poland \\
    \and Key Laboratory for Particle Astrophysics, Institute of High Energy Physics, Chinese Academy of Sciences, 19B Yuquan Road, Beijing 100049, China
     \and Astronomical Observatory of the Jagiellonian University, Orla 171, 30-244 Cracow, Poland\\
    \and Faculty of Physics, University of Bia\l ystok, Cio\l kowskiego 1L, 15-245 Bia\l ystok, Poland \\
     }

\abstract
{The black hole mass measurement in active galaxies is a challenge, particularly in sources where the reverberation method cannot be applied.}
{We aim to determine the black hole mass in a very special object, RE J1034+396, one of the two active galactic nuclei (AGN) with QPO oscillations detected in X-rays, and a single bright AGN with optical band totally dominated by starlight.}
{We fit the stellar content using the code {\sc starlight}, and the broad band disk contribution to optical/UV/X-ray emission is modeled with {\sc optxagnf}. Based on {\sc starlight}, we develop our own code {\sc optgal} for simultaneous fitting of the stellar, Fe II, and BC content in the optical/UV/X-ray data. We also determine the black hole mass using several other independent methods.}  
{Various methods give contradictory results. Most measurements of the black hole mass are in the range  $10^6 - 10^7 M_{\odot}$, and the measurements based on dynamics give higher values than measurements based on H$\beta$ and Mg II emission lines.}
{}

\authorrunning{Czerny et al.}
\titlerunning{Black hole mass in RE J1034+396}
\maketitle

\section{Introduction}

The measurement of the black hole mass in centers of active galaxies is extremely important for a number of reasons. The studies of individual objects  benefit greatly from a black hole mass determination since it is widely accepted that the ratio of the luminosity to the Eddington luminosity is a key parameter that determines the properties of the type-1 unobscured active galactic nuclei (AGN). Eddington ratio is considered to be the leading parameter in the Eigenvector 1, which is determined on the basis of the principal component analysis of the optical spectra and broadband spectral shape (Boroson \& Green 1992;  Sulentic et al. 2000, Marziani et al. 2001; Kuraszkiewicz et al. 2009; Shen \& Ho 2014). Tests of this
hypothesis rely on determining the black hole mass. In cosmology, determining  the black hole mass range as a function of redshift puts strong constraints on the galaxy evolution (e.g. Peng et al. 2006;  Shankar 2009;  Dubois, Volonteri \& Silk 2014,  Aversa et al. 2015). 

Several methods of black hole mass measurements in radio-quiet AGN have been developed (for a reviews, see e.g. Czerny \& Nikolajuk 2010):
\begin{itemize}\itemsep2pt
\item reverberation measurement of the broad line region (BLR) 
\item single spectrum BLR measurement
\item stellar dispersion 
\item narrow line region (NLR) line width
\item bulge luminosity
\item broadband continuum fitting
\item high-frequency break of the X-ray power spectrum
\item X-ray excess variance
\item quasi-periodic oscillations (QPO) in X-ray band
\end{itemize}
We have skipped methods like water maser or binary black hole, since they apply to a few known special sources, and cannot  apply to RE~J1034+396. 

The best established method is based on reverberation studies in the optical band. The measurement of the time delay between the variable continuum emitted by central parts of an accretion disk surrounding a black hole and the response of the  broad emission lines originating in the BLR enables to measure the distance, and the line spectral shape gives the estimate of the orbital velocities of the BLR clouds (e.g. Peterson 1993). This method has been directly applied  to over 50 objects so far (Wandel et al. 1999; Kaspi et al. 2000;  Peterson et al. 2004; Bentz et al. 2009a; Bentz et al. 2010; Denney \& Peterson 2010; Grier et al. 2012; Rafter et al.  2013; Peterson et al. 2014; Du et al. 2014; Du et al. 2015; Hu et al. 2015, Edelson et al. 2015), and some campaigns are under way (King et al. 2015; De Rosa et al. 2015, Shen et al. 2015, Modzelewska et al. 2014, Hryniewicz et al. 2014, Valenti et al. 2015). The method is, however, very demanding in terms of  telescope time.

The second method requires measuring just a single spectrum with an emission line.
Observationally discovered scaling between the BLR size and monochromatic luminosity (see Bentz et al. 2009 for the most recent parameters)  enabled us to replace the reverberation measured delay with a measurement of the monochromatic luminosity. This enabled us to extend the method to thousands of  AGN (e.g. Vestergaard \& Peterson 2006, Kollmeier et al. 2006; Woo et al. 2008; Shen et al. 2011).

One drawback of both methods is that the accurate mass measurement requires an independent determination of the proportionality constant, which implicitly contains  the geometry of the BLR and its kinematics (Collin et al. 2006). This scaling is usually done using the stellar dispersion method for selected sources and varies among the papers (see for example, Section 5 in Bentz et al. 2014). 

The line emission from the gas can also be used for the black hole mass measurement. The simplest variant is to use the width of [O III] line from the NLR as a proxy for the stellar velocity dispersion in the bulge (e.g. Gaskell 2009). However, the emission of the NLR does not have to coincide with the bulge. In the case of the Seyfert galaxy NGC 5548, the variability revealed that most of the [O III] emission comes from the inner 1 - 3 pc (Peterson et al. 2013). 

The stellar dispersion method is the oldest one, but in the context on non-active galaxies. Magorrian et al. (1998) discovered the correlation between the bulge mass and the black hole mass, later followed by an even more tight correlation between the black hole mass and the stellar velocity dispersion (Ferrarese \& Merrit 2000). It was later shown that the same relation applies to AGN, although the stellar velocity dispersion in AGN is not easy to measure (Onken et al. 2004, Grier et al. 2013).
If there is no measurement of any of the quantities mentioned above it is possible to use the bulge luminosity for a proxy of the stellar dispersion, and the recent scaling between the black hole mass and the bulge luminosity is given in Bentz et al. (2009b). 

In general, the mass determination method, based on gas and stellar dynamics, is safe to use  either when the nucleus is resolved and we measure the dynamics within the sphere of the black hole influence, or when the measured quantities represent  the entire bulge well. This second possibility can be easily met for elliptical galaxies but, in the case of spiral galaxies, the measurement contains both the bugle and the disk contribution and any scaling laws may be misleading.

The continuum fitting method is occasionally used for AGN, but it is not simple because of the problems with  accretion disk models and with the data gap between the far-UV and X-rays. Some objects can be well fitted by a standard disk (e.g. Czerny et al. 2011, Capellupo et al. 2015) using only optical/UV spectra. Broadband fits, including X-rays, require a separate description of the X-ray emission as well. The X-ray continuum cannot be well described by the emission of the standard optically thick accretion disk, and the presence of an additional X-ray emitting Comptonizing coronal region is necessary (e.g. Czerny et al. 2003, Done et al. 2012 and the references therein). 

A different family of black hole mass determination methods is based on the X-ray variability. Since we do not have a full understanding of the geometry and the dynamics of the X-ray emitting region (e.g. Edelson et al. 2015, Fabian et al. 2015) this is again a phenomenological approach, justified by studies of sources with known masses. The general shape of the X-ray power spectrum density (PSD) of AGN and galactic sources is, overall, similar, apart from the scaling, showing a steep high frequency tail, a high frequency break and, occasionally, a quasi-periodic oscillation (QPO). The second (low frequency) break is hard to measure in AGN and may not always be present. One black hole mass measurement method uses the dependence of the high frequency break on the black hole mass and the Eddington ratio (McHardy et al. 2006), another one used the normalization of the high frequency tail (Hayashida et al. 1998; Czerny et al. 2001). The dependence of the QPO on the black hole mass was given by Remillard \& McClintock (2006) for the galactic sources.

The reverberation method seems, in general, the most reliable and should be used whenever possible.
However, the reverberation method does not always apply since it means that we must measure well one of the important BLR lines (H$\beta$, Mg II or CIV) and the variable monochromatic continuum. In some cases this is not possible, most frequently because of too high obscuration of the nucleus. Another possible reason is the lack of the observed variability of the optical continuum caused by strong contamination by the stellar emission, and  RE J1034+396 is one such example. The value of the black hole mass in this source is extremely important because this is the first, and the best, example of the QPO discovered in an AGN in the X-ray band (Gierlinski et al. 2008; Alston et al. 2014).

RE~J1034+396 ($z = 0.042443$, after NED\footnote{NASA/IPAC Extragalactic Database}) is an exceptional active galaxy in many aspects. The source has been already contained in the catalog of galaxies and of clusters of galaxies prepared by Zwicky \& Herzog (1966). Initially described as a compact galaxy (Zwicky \& Herzog 1966), it is currently classified as a Narrow Line Seyfert 1 galaxy. It is EUV-bright, and the Big Blue Bump of this Seyfert galaxy has an exceptionally high temperature so the high energy turnover is observed in the soft X-rays (Puchnarewicz et al. 1995). This unique  Big Blue Bump has been subsequently studied in a number of papers (Puchnarewicz et al. 1998, Mason et al. 1996, Wang \& Netzer 2003, Crummy et al. 2006, Casebeer et al. 2006, Done et al. 2012). It is also the first source that unquestionably shows the QPO  in the X-ray emission with the period of $2.7 \times 10^{-4}$ Hz 
(Gierlinski et al. 2008, Middleton, Uttley \& Done
2011; Alston et al. 2014). The warm absorber in this source is not strong, and varies with the QPO phase (Maitra \& Miller 2010).

RE~J1034+396 shows well-developed BLR lines (Puchnarewicz et al. 1995) but the nature of the optical continuum is unclear, and this continuum does not vary (Puchnarewicz et al. 1998). The possibility that the optical/UV continuum comes from a strongly irradiated accretion disk was discussed by Soria \& Puchnarewicz (2002) and  Loska et al. (2004). On the other hand, Bian \& Huang (2010) argue that most of the emission is simply due to the starlight. In modeling, they allowed for the presence of an additional power law but the derived slope ($F_{\lambda} \approx \lambda^{-0.5}$) was not consistent with the tail of an accretion disk. Therefore the source is not a good candidate for reverberation monitoring, and the mean-spectrum approach  is also difficult since it requires  the accretion disk contribution to the continuum to be precisely determined. The soft X-ray emission dominated by the Comptonized disk emission is also mostly constant, and the observed QPO is connected with the variable hard X-ray power-law tail (Middleton et al. 2009).  

 The black hole mass in this galaxy has been estimated by a number of authors using different methods, but the results span a broad range of values (e.g. $(2 - 10) \times 10^6 M_{\odot}$, Puchnarewicz et al. 2001, $(0.6 - 3)\times 10^6 M_{\odot}$, Soria \& Puchnarewicz 2002, and $6.3 \times 10^5 M_{\odot}$, Loska et al. 2004) from broadband continuum models; $(1 - 4) \times 10^6 M_{\odot}$ from stellar velocity dispersion and H$\beta$ line profile, Bian \& Huang 2010; $6.3 \times 10^5, 3.6 \times 10^7$, and $1.3 \times 10^7 M_{\odot}$ from H$\beta$ line width, [O III] line width, and soft X-ray luminosity, correspondingly (Bian \& Zhao 2004). The observed QPO period of $2.7\times 10^{-4}$ Hz (Gierlinski et al. 2008), if identified with the LF QPO dives the black hole mass below 
$4 \times 10^5 M_{\odot}$, corresponding to the Eddington ratio above 10 for the observed bolometric luminosity of 
$5 \times 10^{44}$ ers s$^{-1}$. Identification of the QPO with high frequency oscillations led to values of $6.9 \times 10^6$ or $1.0\times 10^7 M_{\odot}$, depending on the choice of the higher or lower value of the resonance frequency. 

In this paper, we use the available optical/UV and X-ray data and attempt to determine the black hole mass in this object using several of the complementary methods listed above: single spectrum BLR method based on two low ionization lines (LIL) H$\beta$ and Mg II, NLR method, stellar dispersion, broadband fitting of the entire optical/UV/X-ray continuum, with particular attention to the starlight and a QPO method.

\section{Observations}
\label{sect:data}

We collected the archival data from several instruments with the aim to model the broadband spectral energy distribution (SED) covering IR-optical-X-ray band and to apply several independent methods for the black hole mass determination.

In the optical/UV  band we use both the Sloan Digital Sky Survey (SDSS) data and the {\it Hubble Space Telescope} (HST) data obtained with faint object spectrograph (FOS). The SDSS spectrum was taken on 29 December 2003 and HST/FOS data were collected on 31 January 1997. The SDSS data was retrieved through SDSS SkyServer Explorer tool~\footnote{http://cas.sdss.org/dr9/en/tools/explore} from
the DR9 archive. We use data product that has been reduced and calibrated through the standard SDSS automatic pipelines. The HST/FOS data was obtained with the use of the MAST service~\footnote{http://archive.stsci.edu/mast.html}.
The calibrated data product was chosen, which contains the final spectrum processed through the standard pipeline.
We combined the two data sets, but we allow for a gray shift between the two data sets if some weak variability is actually present in the source, in the far UV band. 

\begin{figure}
    \centering
 \includegraphics[width=0.95\hsize]{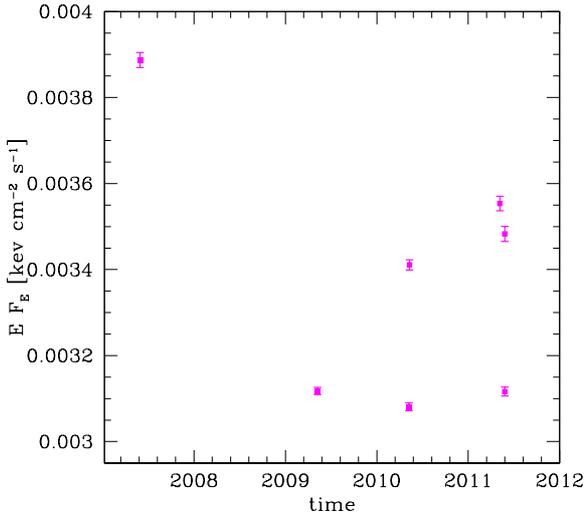}
    \caption{The lightcurve in the far-UV from the OM-XMM instrument at 2952 \AA~ (4.2 eV).} 
    \label{fig:OM_curve}
\end{figure}

These data sets were dereddened for the Galactic extinction with the use of Cardelli et al. extinction law (Cardelli, Clayton \& Mathis 1989). The reddening in this direction is very small ($A_V$ = 0.043 mag based on Galactic
extinction maps from Schlafly \& Finkbeiner (2011) and $R_V$ = 3.1 obtained
through NED, $E_{B-V} = 0.014$) so the reddening correction was not very significant. Since the redshift given in NED was based on the old data set (Karachentsev, Lebedev, \& Shcherbanovskij 1985), we used the [O III]$\lambda5007$ line for better redshift determination. In the further analysis we thus adopt $z =  0.0433 $ for this object. This is only slightly different from the value of 0.043 adopted by Bian \& Huang (2010). 






In X-rays, several measurements  were available from XMM-Newton satellite. The source is clearly variable and we use these measurements to calculate the excess variance and the mean SED in this energy range. We use the 28 lightcurves from the XMM satellite in the 2-10 keV band that were collected in the period from May 2002 to May 2011.  Those lightcurves are mostly short but they are used to calculate the X-ray variance. Finally, we use the longest XMM lightcurves obtained by Middleton et al. (2011) for the broadband spectrum fitting. This 94 ksec observation was performed on 31 May 2007 (OBSID: 0506440101). All X-ray data were extracted from the heasarc archive\footnote{http://heasarc.gsfc.nasa.gov/}.

Since our broadband data (SDSS, HST and long XMM-Newton) come from very different epochs, we allow for an arbitrary additional scaling factor to account for the source variability in the far-UV and X-ray band. We illustrate the flux changes with the OM-XMM monitoring (see Fig.~\ref{fig:OM_curve}). The dimensionless flux dispersion at 2952 \AA~ is 8.8 \%.  The X-ray data come from the period when the source was exceptionally bright in UV (first point in Fig.~\ref{fig:OM_curve}), and the ratio of the the minimum flux to this value is 0.79. The variability in the HST can be higher since the aperture of the OM-XMM contains more starlight than HST. This can be seen from the fact that the flux measured by OM-XMM is almost higher by a factor 2 than the SDSS flux at $\sim 3000$ \AA, and the OM-XMM spectral slope is significantly redder than the SDSS slope, so starlight from a greater distance is also included in OM-XMM.

The integrated optical/UV/X-ray luminosity implies the bolometric luminosity of the source $1.3 \times 10^{44}$ erg s$^{-1}$ cm$^{-2}$ for isotropic emission, and can be, by a factor up to 2, lower if we see the source at low inclination.

\section{Models of the broadband continuum}
\label{sect:models}

The broadband spectrum of a Narrow Line Seyfert galaxy is expected to consist of the accretion disk contribution to the continuum, host galaxy emission (particularly the circumnnuclear stellar cluster), and possibly some dust emission in the red and near-IR. The disk emission is additionally Comptonized in the accretion disk corona responsible for the power-law tail of the spectrum in the X-ray band (Done, Gierli\' nski \& Kubota 2007; Cao 2009; You, Cao \& Yuan 2012). These broad components are supplemented by important localized spectral features connected with AGN activity: BLR lines, Fe II pseudo-continuum, and Balmer continuum. We do not model the BLR lines, but instead we mask the strong emission line regions in the spectrum. However, we model all the remaining spectral components since they are essential for the correct decomposition of the spectrum.

\subsection{Starlight contribution}

As it was shown by Bian \& Huang (2010), the optical spectrum of RE J1034 is mostly dominated by the starlight. However, the traces of the accretion disk contribution are also visible and in Bian \& Huang (2010) they were modeled by a power law of an arbitrary slope. The best-fit slope ($F_{\lambda} \propto \lambda^{-0.58}$, $F_{\nu} \propto \nu^{-1.42}$), however, is clearly inconsistent with the expectations of the accretion disk theory (Shakura \& Sunyaev 1973) ($F_{\lambda} \propto \lambda^{-7/3}$).  Therefore, we repeat the starlight analysis of Bian \& Huang (2010), first using the same 
publicly available STARLIGHT\footnote{http://astro.ufsc.br/starlight/node/1} code (Cid Fernandes et al. 2005, 2009), and then
preparing our own fitting code, {\sc optgal,} which can be used for fitting both opt/UV and X-ray data sets simultaneously. The basic scientific content is the same as that of STARLIGHT. We use the same 45 templates of Bruzual \& Charlot (2003). We allow for the intrinsic reddening of the stellar cluster which is described with the Cardelli et al. (1989) curve. We also include the option of a relative velocity shift of the stellar component with respect to the systemic redshift, and we include the stellar velocity dispersion. More technical details are provided  in Sect.~\ref{sect:fitting}.

\subsection{Fe II pseudo-continuum}
\label{subsection:FeII}

The contribution of the Fe II emission is seen both in the optical and in the UV range. Modeling quasars (Hryniewicz et al. 2014, Modzelewska et al. 2014), we found that the theoretical models of the Fe II emission smeared owing to the velocity dispersion of an order of $\sim 900$ km s$^{-1}$ are actually better than observational templates, particularly in the region of Mg II$\lambda2800$. Each of their templates is calculated for a different value of the density, turbulent velocity, and ionization parameter. As a reference, we use the best-fit template for CTS C30.10 (Modzelewska et al. 2014), corresponding to a local density ($n = 10^{12}$ cm$^{-3}$, turbulent velocity of 20 km s$^{-1}$ and ionization parameter $\Phi = 10^{20.5}$ cm$^{-2}$s$^{-2}$). However, these templates do not cover the shortest wavelengths below 2000 \AA. In this region, observational templates are better since they also contain other Fe contributions. We thus use the models of Bruhweiler \& Verner (2008) above 2250 \AA ~and those of Vestergaard \& Wilkes (2001) below 2250 \AA, in the \verb+Fe_UVtemplt_B+ version. The theoretical templates were broadened with a Gaussian profile, assuming 900 km s$^{-1}$. 

\subsection{Balmer continuum}

To model the Balmer continuum (BC), we repeat the procedure described in Dietrich et al. (2002) and the references therein. Blueward of the Balmer
edge ($\lambda \approx 3675$\,\AA), this feature is described by the Planck function $B_{\nu}(T_{e})$ (Grandi 1982) with constant electron
temperature of 15\,000 K. The optical depth is not assumed to be constant, but its change with wavelength is computed using a simple formula:
\begin{equation}
\tau_{\nu} = \tau_{BE} \left(\frac{\nu}{\nu_{BE}}\right)^{-3},
\end{equation}
where $\tau_{BE}$ is the optical depth at the Balmer edge radiation frequency ($\nu_{BE}$).
Redward of the Balmer edge blend of hydrogen emission lines is generated. This was performed using atomic data provided by Storey \& Hummer
(1995) with the recombination line intensities for case B (opaque nebula), $T_{e}$ = 15\,000 K, $n_{e} = 10^{8}-10^{10}\, {\rm cm}^{-3}$.
The accounted Balmer emission lines covered excitation top level for the transitions in the range $10 \leq n \leq 50$.

\subsection{Accretion disk model of the broadband spectrum}

The X-ray data of RE~J1034+396  explored in the next section was collected by XMM-Newton on 2007-05-31 and 2007-06-01, with total observing time $\sim$ 94 ks. The extensive spectral analysis of this observation was performed by Middleton et al. (2009). They considered a number of models as possible interpretations of the observed continuum: (1) low-temperature Comptonization of the disk emission for the soft excess and power law for high energy tail; (2) the steep power-law illuminating spectrum and the smeared reflection spectra from double reflectors in different ionization states; (3) the smeared absorption of a steep power-law spectrum by double partially-ionized winds; (4) the absorption but by the clumpy partially-ionized winds. The viable model is determined to simultaneously fit both the X-ray continuum and the rapid variability dominated by the QPO. The detailed explorations are described in Sect. 4 of Middleton et al. (2009). The conclusion is that the observed overall spectrum at 0.3-10 keV should be decomposed into the low-temperature Comptonization of the disk emission in the low energy, and the power-law continuum from the high-temperature Comptonization. It does not uniquely sets the geometrical arrangement of the system. 

For the description of the accretion disk emission, we thus use the model {\sc optxagnf} developed by Done
et al. (2012) for the purpose of modeling NLS1, and implemented in XSPEC (Arnaud 1996). This phenomenological model  represents the complexity of the disk emission well. Bare disk in this model is seen for radii larger than $R_{cor}$, and the inner disk region is covered by the disk corona. The local disk emission is color-corrected using the approximations given in Done et al. (2012). Inside $R_{cor}$, part of the energy is dissipated in the corona. Two thermal Comptonization media are necessary. Low-temperature Comptonization is necessary to model the soft X-ray excess. This soft Comptonization is described by the electron temperature, $T_e$, and the optical depth, $\tau$, and implemented in XSPEC as {\sc comptt} (Titarchuk 1994). The second Comptonization has to reproduce the power-law emission which dominates above 2 keV. This high-temperature component is described by {\sc nthcomp} (Zdziarski et al. 1996), and parameterized by the hard X-ray slope, $\Gamma$, and the fraction of energy, $f_{pl}$, dissipated in the hard X-ray emitting corona, while the temperature of the hot coronal phase is fixed at 100 keV. The soft part of the corona may physically correspond to an optically thick skin on the top of the disk, and it is likely strongly magnetized (R\' o\. za\' nska et al. 2015). This model is basically consistent with the lack of relativistically smeared K$\alpha$ line but with the reprocessing in a region of the size of $\sim 150$ s, as measured from the time delays between X-ray bands (Zoghbi \& Fabian 2011), as well as with the existence of the Shakura-Sunyaev disk in AGN at larger radii (e.g. Edelson et al. 2015).

\begin{figure}
    \centering
 \includegraphics[width=0.95\hsize]{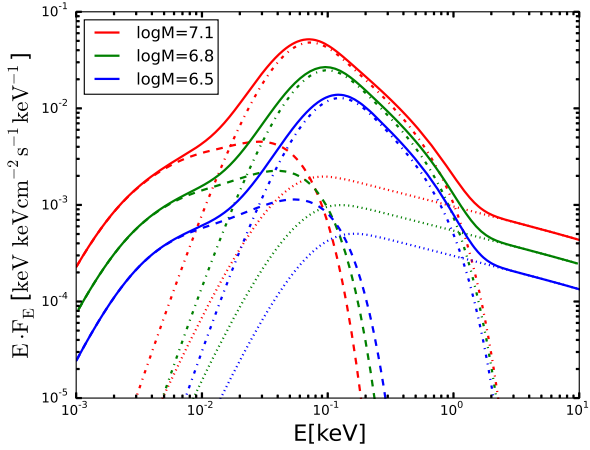}
    \caption{The model spectra of {\sc optxagnf} for $\rm logM_{\rm BH}/M_{\odot}$= 6.5 (blue), 6.8 (green), 7.1 (red). Other parameter values were fixed as in the best fit of Middleton et al. (2011). The dashed, dash-dotted and dotted lines represent the disk blackbody, soft Compton, and hard Compton components, respectively. The total spectra are plotted with the solid lines.}
    \label{fig:model_spectrum}
\end{figure}

Input parameters of the model are the black hole mass, $M_{\rm BH}$, black hole spin, $a$, bolometric Eddington ratio, $L/L_{\rm Edd}$, $R_{cor}$,  $T_e$, $\tau$, $\Gamma$, $f_{pl}$,  the comoving (proper) distance of the source, and redshift. To demonstrate the influence of the mass on the overall spectra, we plot the model spectra of {\sc optxagnf} for the cases of three different black hole masses $\rm log M_{BH} = 5.6, 6.8$ and 7.1 (see Fig.~\ref{fig:model_spectrum}), while the rest of the parameters are not changed.  On the one hand, it can be easily understood that the overall luminosity (the disk, soft Comptonization, and hard Comptonization emission) increases with the mass, as the flux of the seed photons from the disk is set by the combination of $M_{\rm BH}\dot{M} \propto M_{\rm BH}^2L/L_{Edd}$ (e.g Davis \& Laor 2011). On the other hand, according to the standard accretion disk theory, the disk temperature at a given radius decreases with $M_{\rm BH}$, which will reduce the energy of the seed photons of blackbody radiation. Therefore, the disk, soft Comptonization, and hard Comptonization spectra will accordingly be shifted to low energy band. There is no significant variation in the shapes of the three overall spectra, because other model parameters, $L/L_{\rm Edd}$, $R_{cor}$,  $T_e$, $\tau$, $\Gamma$, $f_{pl}$ are identical, respectively.

\section{Continuum fitting methods}
\label{sect:fitting}

The model contains numerous free parameters,  45 of these are just normalizations of different starlight components in the STARLIGHT code, and the X-ray data model adopted after Middleton et al. (2011) also has  its complexity. We divided the fitting procedure into two parts.

We first model the optical/UV part alone, in the rest frame. This approach also has  the advantage of giving results directly comparable to Bian \& Huang (2010). The optical/UV data were thus corrected for the Galactic extinction, moved to the rest frame, and rebinned to the 1 \AA~ bins as requested by the STARLIGHT code.

In that part we (temporarily) replace the disk component with the power law, as in Bian \& Huang (2010). We subtract Fe II emission and Balmer continuum from the data, using a dense grid of the normalizations of these components, and then we run the  STARLIGHT code with a pre-set power-law slope to get the starlight shape given by 45 normalizations of the components, the systematic stellar velocity, stellar velocity dispersion, and the reddening. The resulting $\chi^2$  is used to select the best-fit solution. Regions of  strong BLR emissions are masked and do not contribute to the fit quality.  The gap between the SDSS and HST data is masked as well, and the shift in the HST data is allowed.

Next we perform the global fit of the combined optical/UV/X-ray data. This is performed in the observed frame using the original optical and UV data without rebinning, but with dereddening, and the X-ray data from XMM. For this purpose, we created a unique new XSPEC model, which is based on the conceptual content of STARLIGHT, but without their fitting procedure. It also includes the pseudo-continua needed to fit the data (Fe II and BC). The new model, {\sc optgal}, as a subroutine of XSPEC, written in Fortran, provides the combined emission of all stellar components, Fe II, and BC for assumed model parameters. It thus describes the optical/UV spectrum of an AGN up to 1 500 \AA~ but with the aim of using  it together with the X-ray data within the XSPEC.  Our model contains 56 parameters and we believe it is one of the largest X-spec models created. The parameters are: 45 normalizations of the stellar components of corresponding ages and metallicity, normalization of the BC, Fe II components (4), the stellar component shift, stellar component dispersion, local star cluster extinction, source redshift, and overall normalization.

To represent the broadband optical/UV/X-ray disk component, we use the model {\sc optxagnf} developed by Done et al. {2011} and found by Middleton et al. (2011) to best represent  the XMM data set we use.
The final XSPEC model used to fit the broad band data is thus {\sc optgal +tbabs*optxagnf}.

The optical and UV data are expressed in units of keV cm$^{-2}$s$^{-1}$ to combine with the X-ray data. The data regions of strong emission lines are masked in exactly the same way as in the STARLIGHT code, and the relative shift between the SDSS and HST data is allowed since they do not come from the same epoch.
The search for a best fit solution is done by XSPEC by minimizing the $\chi^2$.

\section{Results of the continuum fitting}

Following the procedure described in Sect.~\ref{sect:fitting} we present here the results of the fitting of all components: starlight, BC, Fe II, and accretion disk to the broadband optical/UV/X-ray data.
We emphasise that here we adopt the value of redshift 0.0433 for this source, as determined from the [O III]$\lambda$5007 \AA~ line. This value slightly differs from the value given by NED (z = 0.042443) and the value of z = 0.043 adopted by Bian \& Huang (2010). The results for optical/UV fitting of all components are given in Table~\ref{table:FirstStage}, and the results for disk-component fitting are given in Table~\ref{tab:optagn_fitting}.

\subsection{Opt/UV fitting}
\label{sect:opt/uv_fitting}

The fitting of the optical/UV continuum alone is interesting in itself since it provides a direct comparison with the previous results obtained by Bian \& Huang (2010), who fitted only the optical SDSS data.

\begin{table*}
\caption{STARLIGHT fitting results for a disk contribution described as a power law with a fixed slope (upper part).}   
\label{table:FirstStage}      
\centering                          
\begin{tabular}{l c c  c c r r r}        
\hline\hline      
slope    & Fe II   &   BC    & $f_{\rm HST/SDSS}$ &  $A_V$   &   $v_d$  &  $\chi^2/dof$ \\
         & [$10^{-8}$]     &   &         &          &   km s$^{-1}$ & \\ 
1 & 2 & 3 & 4 & 5 & 6 & 7  \\
\hline
-0.5     &    1.9  &     -      &  -        & 0.004   & 0.05 &  1.25  \\       
-7/3     &    2.8  &     -      &  -        & 0.60    &  72  &  1.55 \\
-7/3     &    1.9  &   0.6      &  -        & 0.60    &  73  &  1.51 \\
-7/3     &    1.6  &   1.2      & 0.78      & 0.38    & 124  &  1.40  \\
\hline
\end{tabular}
\\[0.5ex]
First column shows the assumed fixed power-law component representing the disk emission; second column is the normalization of the Fe II pseudo-continuum; third column is the normalization of the Balmer continuum; column 4 gives the scaling factor between the HST and SDSS data; column 5 gives the intrinsic extinction for the stellar cluster; column 6 contains the stellar velocity dispersion and the last column gives the reduced $\chi^2$ for one degree of freedom. The normalizations of BC and Fe II components do not have the direct physical meaning, they normalize the corresponding templates.
\end{table*}

We first searched for the best solution to the optical/UV spectra without a Balmer continuum contribution, and without any constraints for the slope of the underlying power law. The best-fit solution was quite close to the solution found by Bian \& Huang (2010). The implied slope was $\alpha_{PL} = -0.5$ (in $F_{\lambda} \propto \lambda^{\alpha}$ convention) although we used the HST data as well, and we fitted the spectral range from 1500 \AA ~ to 8900 \AA.  The fit quality
is actually better than in Bian \& Huang (2010), with $\chi^2/dof = 1.25$. The stellar systematic velocity is 133 km s$^{-1}$, corresponding to a spectral shift by 2.2 \AA~
to the red. If interpreted as a redshift inaccuracy, it would imply a shift by 0.0004. Determination of the redshift from [O III]$\lambda 5007$ seems more accurate, 
so this shift is likely real. The derived velocity dispersion is  consistent with zero, i.e. the broadening of the features intrinsic to the models is high enough to match the data and the intrinsic dispersion is not measurable. 
Since the average instrumental resolution of HST/FOS data for this object is 56 km/s, the SDSS resolution is 60 km/s (Greene \& Ho 2005), and the resolution of the templates is 86 km s$^{-1}$ (Cid Fernandes et al. 2005). Using Eq.~3 of Bian \& Huang (2010), we thus have an upper limit for the stellar velocity dispersion of 65 km s$^{-1}$.  In this case the stellar cluster does not require significant reddening ($A_V = 0.0004$). The stellar population results are similar to Bian \& Huang (2010) results, strong contribution from the young stars of $\sim 5 \times 10^6$ yr is very important. The disk modeled by the power law contributes 63\% of the light at 4020~ \AA.

However, the best-fit slope derived in the course of fitting is inconsistent with the disk contribution. Therefore, we also consider a case where the power-law slope is frozen at 
$\alpha = -7/3$, as in Shakura-Sunyaev disk. This type of fit is formally worse. The best  fit has $\chi^2/dof = 1.55$. The stellar cluster systematic velocity is similar to the previous case (134 km s$^{-1}$) but the velocity dispersion required by the model is higher,  73 km s$^{-1}$ . Now the cluster requires significant reddening ($A_V = 0.60$). The intrinsic stellar velocity obtained from the steep power-law solution is much higher, 95 km s$^{-1}$. The power-law contribution in this solution is much lower, only 23\% at 4020 \AA. 

Bian \& Huang (2010) analyzed the stellar content of the circumnuclear stellar cluster in great detail. Our first solution was very similar to theirs. The second solution is not much different. There are clearly two populations of stars: old, metal poor stars (Z = 0.004), ages of  5 - 10 Gyr, and young high metallicity stars (Z = 0.02 - 0.05), with ages in the range of  3 - 100 Myr. The dominating stars are somewhat older than in the previous case, with age $\sim 10^7$ yr.  

\subsection{Broadband fitting to opt/UV/X-ray data}

We now combine the opt/UV spectrum from SDSS and HST in the observed frame with the long XMM observation and perform
the global fitting using our new XSPEC model {\sc optgal} that models the starlight and the pseudo-continua Fe II and BC, and {\sc optxagnf} from XSPEC package (Arnaud 1996). 
The disk model {\sc optxagnf} is described in detail in Sect. 3.4. Our base model to fit the broadband continuum reads {\sc optgal + tbabs*optxagnf}, in which the Galactic value $N_{\rm H}$ of {\sc tbabs} is $1.47 \times 10^{20}$ $\rm cm^{-2}$ (from the $n_H$ tool in
High Energy Astrophysics Archive Research Center (HEASARC)\footnote{http://heasarc.gsfc.nasa.gov/cgi-bin/Tools/w3nh/w3nh.pl}, consistently with Middleton et al. 2011). The input parameters of the {\sc optxagnf} model for the disk components are the black hole mass, $ 1 M_{\odot}\le M_{\rm BH} \le 10^9M_{\odot}$, black hole spin, $0.0 \le a \le 0.998$, bolometric Eddington ratio, $-10 \le \rm log$ $L/L_{\rm Edd} \le 2$ (formal limits), coronal radius $1 \le r_{\rm cor} \le 100$ in units of $r_g = G M/c^2$, the outer radius of the disk $3\le \rm log$ $r_{\rm out}\le 7$. The inclination angle in the {\sc optxagnf} model is fixed with $60^{\circ}$, although the inclination angle could change the normalization by a factor $\sim$ 2. Moreover, both the soft and hard Comptonization, i.e. electron temperature $T_{\rm e}$, optical depth of the soft Comptonization component $\tau$, and spectral index of the hard Comptonization component $\Gamma$  are fixed with the values derived from the X-ray spectrum fitting alone by Middleton et al. (2011), except that the input parameter, $0 \le f_{\rm pl} \le 1$ is a free parameter in our fitting, which represents the fraction of the energy emitted in the hard Comptonization component. Altogether, our combined model has  72 parameters representing the normalizations of the stellar components and all other elements as described in Sect~\ref{sect:models}, as well as the parameters built into the disk/corona model, which best fits  the XMM-Newton data alone.
We allow for a variable factor scale between the SDSS and HST data, but we do not introduce any scaling between SDSS and XMM-Newton in our basic fits. We address this point later.

The search for  a global minimum with such a complex model is difficult and cannot be done fully automatically. We performed the search for the best solution using a constant step grid for parameters of interests (the steppar option in XSPEC), in particular for the black hole mass, and the errors were derived assuming that the small (close to zero) parameters in the starlight components are fixed at zero level in the contour error search.

\begin{figure}
    \centering
 \includegraphics[width=0.95\hsize]{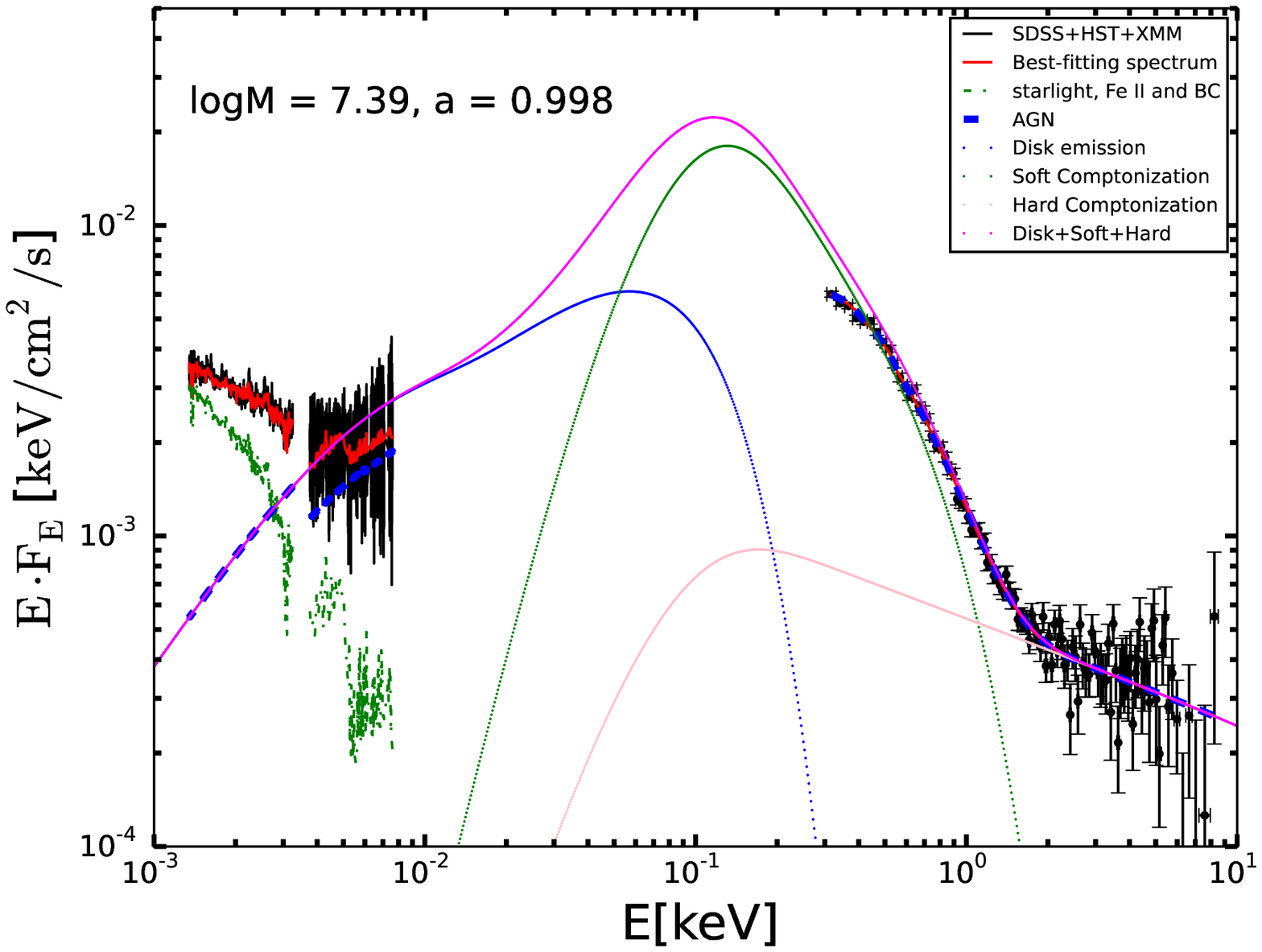}
 \includegraphics[width=0.95\hsize]{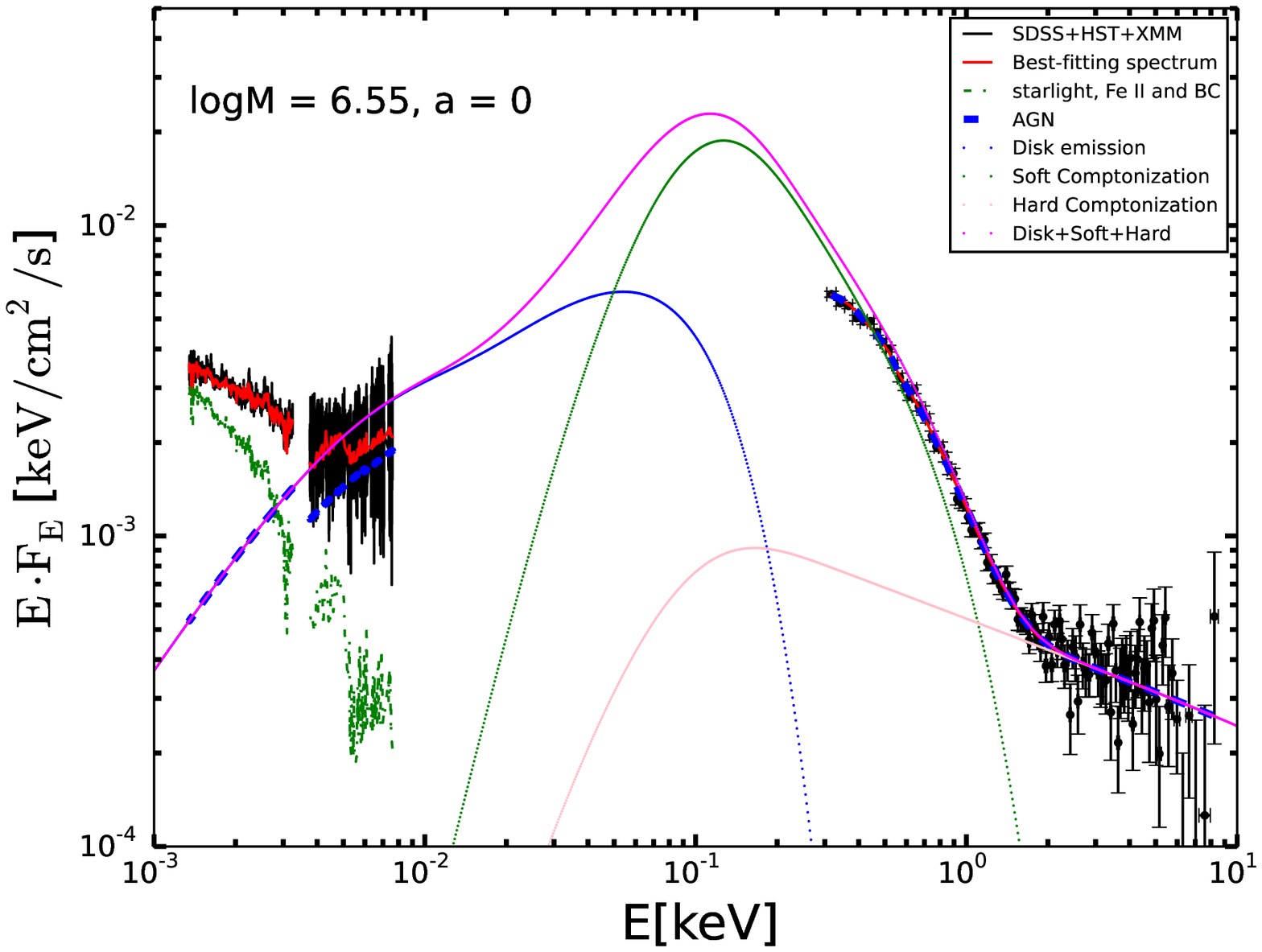}
    \caption{The broadband fitting of the disk to the optical/UV/X-ray data for a = 0.998 (upper panel, best fit $\rm logM_{\rm BH}/M_{\odot}= 7.39$) and a = 0 (lower panel,  best fit $\rm logM_{\rm BH}/M_{\odot}= 6.55$). The red solid line represents the models {\sc optxagnf} with the best-fitting values given in Table 2, and the Galactic extinction {\sc tbabs} and the HST data shift applied to the data  is not included here, which explains the departure of the data/model in soft X-rays and HST disk plot. The blue, green, 
and pink dashed lines represent the disk blackbody, soft Compton, and hard Compton components, respectively.}
    \label{fig:optagnfit2}
\end{figure}

We cannot fit all the parameters uniquely, so we consider in detail two solutions for a fixed black hole spin: $a = 0$ and $a = 0.998$. The results are summarized in Table~\ref{tab:optagn_fitting} and plotted in 
Fig.~\ref{fig:optagnfit2}. 

Better fit was obtained in the case of the maximally rotating black hole. The black hole mass obtained is relatively large, $M_{BH} = (2.47 \pm 0.15) \times 10^7 M_{\odot}$, the Eddington ratio is rather low, and the corona covers only a relatively small part of the disk ($r_{cor}= 7.3 r_g$). Still, the effect of Comptonization is strong, i.e. the fraction of the energy emitted by the hot corona is 67\%. This is because the emissivity in a Novikov-Thorne disk (Novikov \& Thorne 1973) around a fast rotating black hole is strongly concentrated towards the black hole. The {\sc optxagnf} model used to describe the disk/corona emission follows the dissipation profile of the standard accretion disk to control  the accretion energetics well.

Fit for the case of a non-rotating black hole is formally worse but the difference in $\chi^2$ is only by 10, for 6278 d.o.f. (degrees of freedom). Taking into account the model complexity, we cannot favor any of the two solutions. The black hole mass in this solution is much lower, $M_{BH} = 3.54^{+0.25}_{-0.45} \times 10^6 M_{\odot}$, the Eddington ratio is higher, close to 1, and the corona is very extended  ($r_{cor}= 54.8 r_g$). The solutions look very similar because the maximum of the disk emission is mostly determined by the temperature of the corona, fixed to be the same in both cases. The fraction of the energy dissipated in the corona is again 67 \%, the same within the uncertainty as in the previous case, despite much larger corona size since the Novikov-Thorne disk emissivity for a non-rotating black hole peaks at a much larger distance. The similarity of these two solutions is simply required by the data points that strongly constrain the fit both at low and at high energies. Also the stellar parameters derived from those two solutions are very similar, including stellar dispersion and extinction.

Fits are not perfect since the optical/UV spectrum is rich in details that are not fully modeled by the available starlight and Fe II templates. Also some of the fainter emission lines are not properly masked. Taking this into account, we consider the fit's quality as basically satisfactory.

These two extreme mass values give the range where the broad band continuum model can be well fit to the available data. For every value of the spin, we expect to find a corresponding value of the black hole mass. We did not make these computations since the fitting is very time-consuming, and it would not provide a unique solution that fixes both the black hole mass and the spin. 

We also performed fits for the three black hole masses given by Gierlinski et al. 2008, but in the case of a fixed maximally rotating black hole.  The results are given in Table~\ref{tab:optagn_fitting}. Fits are worse than before. In the case of the first two black hole mass values, a better solution may be found if the spin is treated as a free parameter, and lower spin would be clearly favored. On the other hand, the solutions are again quite similar, and together they form a sequence of increasing black hole mass and decreasing the Eddington ratio. The optical luminosity $N \propto (M_{\rm BH}\dot{M})^{2/3}=[M_{\rm BH}\cdot 10^{\rm logL/L_{\rm Edd}}\cdot L_{\rm Edd}/(\eta c^2)]^{2/3}$ (Davis \& Laor 2011), where $L_{\rm Edd}=1.26 \times 10^{38}M_{\rm BH}/M_{\odot}\ \rm erg/s$, the efficiency $\eta$ can be derived as a function of the spin (Novikov \& Thorne 1973; You et al 2015).  The smallest value of the mass cannot be fitted properly even if we assume a non-rotating black hole. This black hole mass is lower than the mass obtained with assumed $a = 0$. The implied Eddington ratio is larger than 1 but it is not enough to  model the continuum level accurately in the optical/UV band. This is partially compensated for by the decrease in the extinction in the starlight contribution and a relative enhancement in the number of younger stars by a factor of 2. However, this additional starlight contribution does not represent the overall shape of the spectrum in the UV part as the solutions with a stronger contribution from the disk, characteristic for cases with larger masses. 

The shift between HST and SDSS data does not depend on the black hole mass. Roughly the same value was requested by a power law fit to the opt/UV data alone. Otherwise, Balmer Continuum is not required by the data, while its presence is usually expected. The fits improves when the shift between the HST and SDSS data are both allowed. The value ($\sim 0.69$) is somewhat lower than the measured extreme ratio in the OM-XMM data (0.79; see Fig.~\ref{fig:OM_curve}), but as we argued in Sect.~\ref{sect:data}, the expected amplitude is higher in HST than in OM-XM, because of lower starlight contamination in HST.

Since the X-ray emission also shows the variability, we consider the two cases separately: X-ray emission higher by a factor of 2 and X-ray emission lower by a factor of 2 than in the data set used by us. Solutions with intermediate black hole mass, $6.9\times 10^6 M_{\odot}$, give $\chi^2/\rm dof $ 1.54, and 1.61. In the second case, the fit is worse since, as before, starlight alone is not a good representation of the UV data. We cannot obtain a fit with the normalization of the XMM-Newton to SDSS data as a free parameters owing to the problems with model convergence. Non-linear coupling and degeneracy between the data normalization, extinction, and the fractions of younger stars in the population prevent a successful automatic solution within XSPEC.

We did not calculate the full contour errors for the other parameters of the model apart from those given in Table~\ref{tab:optagn_fitting}, and for the stellar vellocity dispersion that is later used for the mass determination. However, comparing the values obtained from the two solutions for $a = 0$ and $a = 0.998,$ we can clearly see  how accurately they can be determined, independently from other parameters. In both cases, the dominant contribution to the starlight comes from the same types of stars. The ratio of the stars older than $10^9$ years to the stars younger than $10^9$ years is 0.89 and 0.92, correspondingly. The systemic shift between the AGN reference frame based on [O III] line ($ z = 0.0433$) and the starlight is -126.3 km s$^{-1}$ and -127.7 km s$^{-1}$ in the two cases. The contribution of the starlight to the optical emission at 4020 \AA~ is 61\% and 62\%, correspondingly. Some of those values, however, depend on the description of the disk contribution, as it can be seen from a comparison of Table~1 and Table~2.

\begin{table*}[ht!]
\caption{Selected fits of {\sc optgal + tbabs*optxagnf} to broad band continuum of RE J1034+396 for two values of black hole spin.}
\centering
\begin{tabular}{cccccccccc}
\hline \hline \\  [1pt]
$\rm log\ M_{\rm BH}/M_{\odot}$ & 
$\rm log$ $(L/L_{\rm Edd})$ & 
$a$ &
$r_{\rm cor}$ &
$f_{\rm pl}$ [\%] &
$Fe II$ &
$BC$ &
$A_V$ &
$f_{\rm HST/SDSS}$ &
$\chi^2$/dof ($\chi^2_{\nu}$) \\[5pt] 
\hline \\
$7.39^{+0.03}_{0.03}$ & 
$-1.03^{+0.05}_{-0.03}$ &
$0.998$ &
$7.3^{+3.0}_{-1.4}$ &
$10.2^{+1.6}_{-1.8}$ &
$2.67^{+0.25}_{-0.23}$ &
$0.092^{+0.009}_{-0.005}$ &
$0.79^{+0.004}_{-0.004}$ &
$0.690^{+0.009}_{-0.0045}$ &
9549.82/6278 (1.52) \\ [5pt]
\hline \\
$6.55^{+0.03}_{-0.06}$ & 
$-0.14^{+0.09}_{-0.03}$ &
$0.0$ &
$54.8^{+18.7}_{-13.0}$ &
$9.9^{+1.9}_{-1.0}$ &
$2.61^{+0.09}_{-0.17}$ &
$0.092^{+0.006}_{-0.006}$ &
$0.81^{+0.004}_{-0.004}$ &
$0.689^{+0.010}_{-0.009}$ &
9559.12/6278 (1.52) \\ [5pt]
\hline 
\hline \\
%
%
7.0 & 
$-0.41^{+0.01}_{-0.02}$ &
$0.998$ &
$34.1^{+1.2}_{-1.4}$ &
$5.8^{+0.2}_{-0.2}$ &
$2.34^{+0.05}_{-0.13}$ &
$0.115^{+0.008}_{-0.007}$ &
$0.75^{+0.04}_{-0.03}$ &
$0.710^{+0.012}_{-0.010}$ &
9676.73/6279 (1.54) \\ [5pt]
\hline \\
6.8 & 
$-0.217^{+0.006}_{-0.011}$ &
$0.998$ &
$49.2^{+0.3}_{-1.4}$ &
$5.4^{+0.2}_{-0.2}$ &
$2.17^{+0.13}_{-0.10}$ &
$0.128^{+0.004}_{-0.005}$ &
$0.72^{+0.02}_{-0.04}$ &
$0.726^{+0.012}_{-0.011}$ &
9829.86/6279 (1.57) \\ [5pt]
\hline \\

5.6 & 
$0.538^{+0.004}_{-0.023}$ &
$0.998$ &
100 &
$10.9^{+0.4}_{-0.3}$ &
$2.56^{+0.09}_{-0.10}$ &
$0.170^{+0.003}_{-0.006}$ &
$0.30^{+0.02}_{-0.02}$ &
$ 0.707^{+0.005}_{-0.004}$ &
10596.41/6279 (1.69) \\ [5pt]

\hline
\end{tabular}
\tablefoot{The best-fitting spectral parameters. All errors are quoted at the $90 \%$ confidence level ($\Delta \chi^2 = 2.706$). $\rm M_{\rm BH}$ is in units of $M_{\odot}$. $r_{\rm cor}$ is in units of $r_g=GM_{\rm BH}/c^2$, and $\rm log$ $r_{\rm out}$ (in the same units) is fixed at 5.0. The two upper lines give the black hole mass obtained from the fitting, the three lower lines are fits with the black hole mass fixed at values suggested by QPO (Gierlinski et al. 2008). Other model parameters were frozen at the values favored by Middleton et al. (2011): $kT_e = 0.195$ keV, $\tau = 14.1$, $\Gamma = 2.33$. The normalization of Fe II is in units of $10^{-8}$, as in Table~\ref{table:FirstStage}. Value of $f_{\rm HST/SDSS}$  is fitted as a constant, accounting for the SDSS and HST data sets coming from different epochs.} 
\label{tab:optagn_fitting}
\end{table*} 
 

\subsection{Spectral decomposition in opt/UV from opt/UV/X-ray fitting}

We chose the solution for the non-rotating black hole (black hole mass $M = 3.54 \times 10^6 M_{\odot}$) for a detailed discussion in this section. The corresponding results of model fitting are plotted in Fig.~\ref{fig:optagnfit2}. The solution for $a = 0.998$ and for the two other large mass cases from Table~2  give very similar results for the starlight properties.

\begin{figure*}
    \centering
\includegraphics[width=0.95 \hsize]{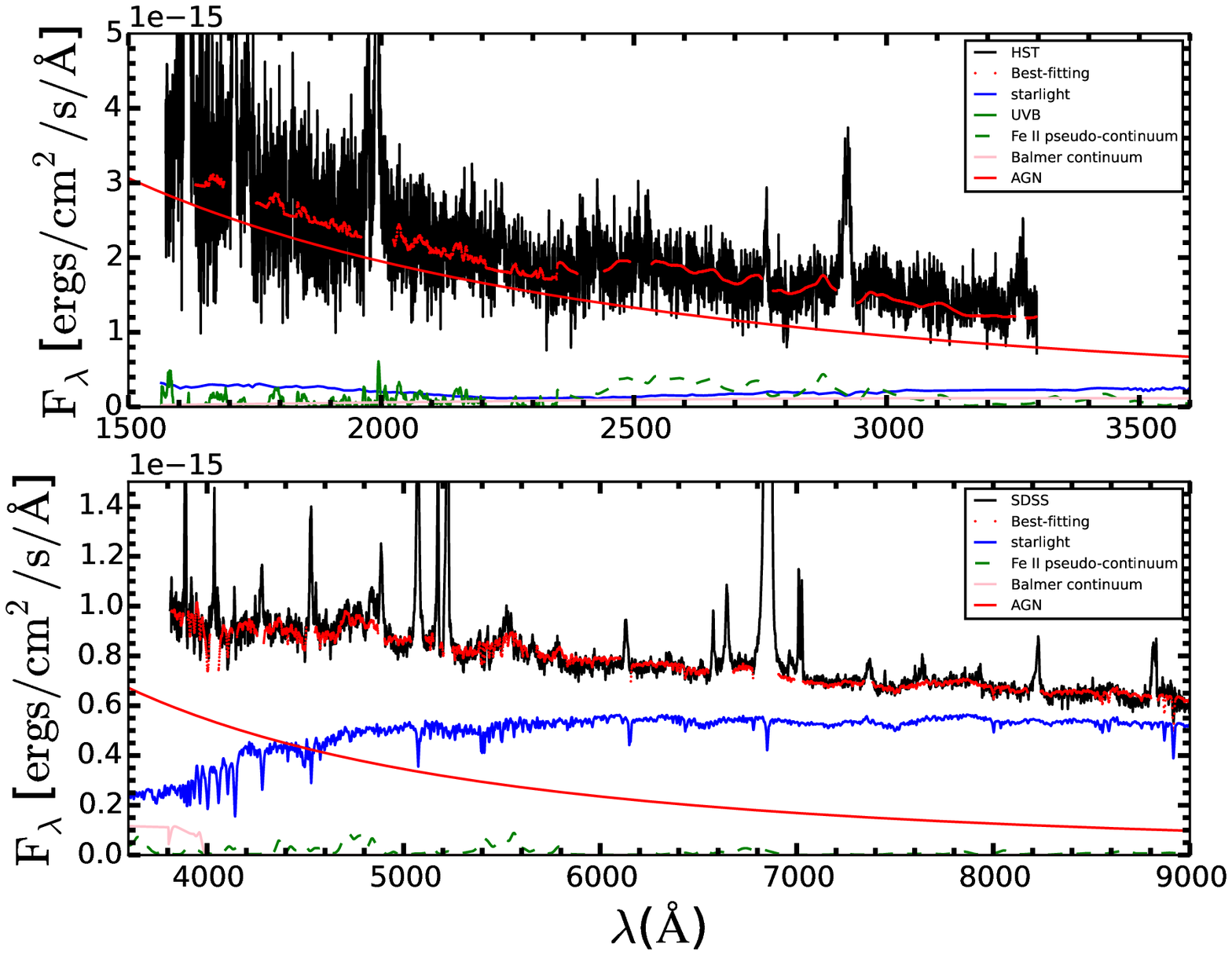}
    \caption{The fit of the $a = 0$ ($\log M = 6.55$) model (see Table~2) to the optical/UV spectrum of RE J1034+396 (red line) to the data (black line);  red line below shows the disk contribution, pink line - Balmer Continuum, green solid line the Fe II contribution below 2250 \AA~(Vestergaard \& Wilkes 2001), green dashed line the Fe II above 2250 \AA~(Bruhweiler \& Verner 2008). Stars contribute significantly to the spectrum  even in UV part. The HST data are renormalized by constant 0.717.}
    \label{fig:opt}
\end{figure*}

\begin{figure}
    \centering
\includegraphics[width=0.95\hsize]{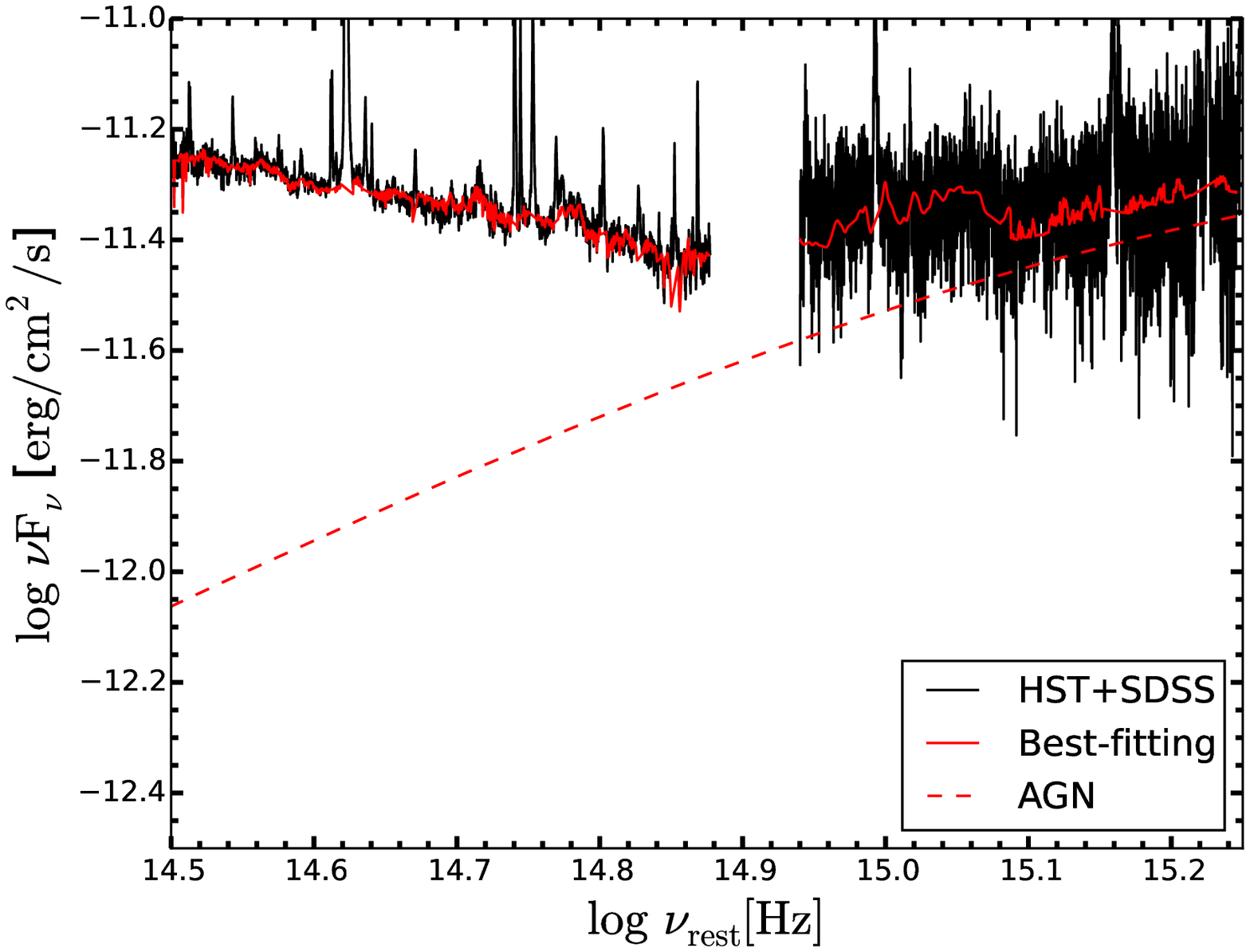}
    \caption{The fit of the $a = 0$ model in the optical/UV range in $\nu F_{\nu}$ (red line) to the data (black line). The disk contribution is marked with the dashed line.}
    \label{fig:optnuFnu}
\end{figure}

The shape of the disk component resulting from broadband fit is clearly different from a power law. This affects the decomposition in the optical/UV band. 
The solution in $F_{\lambda}$ space is shown in Fig.~\ref{fig:opt}. The fit seem satisfactory in the optical band. In the UV part, the Fe II template  probably does not  account for the spectral features well. 
The same result, but in $\nu F_{\nu}$ space is shown in Fig.~\ref{fig:optnuFnu}. Here we also conveniently plot the observationally determined disk contribution.

The disk contribution to the total flux is at the level of 61\% at 4020 \AA, and at the level of still only 88\%  at 2000 \AA. This is consistent with Bian \& Huang (2010), where the disk (a power law) exceeded 50 \% at 4020 \AA. Their analysis did not extend beyond 3400 \AA.

\begin{figure}
    \centering
 \includegraphics[width=0.95\hsize]{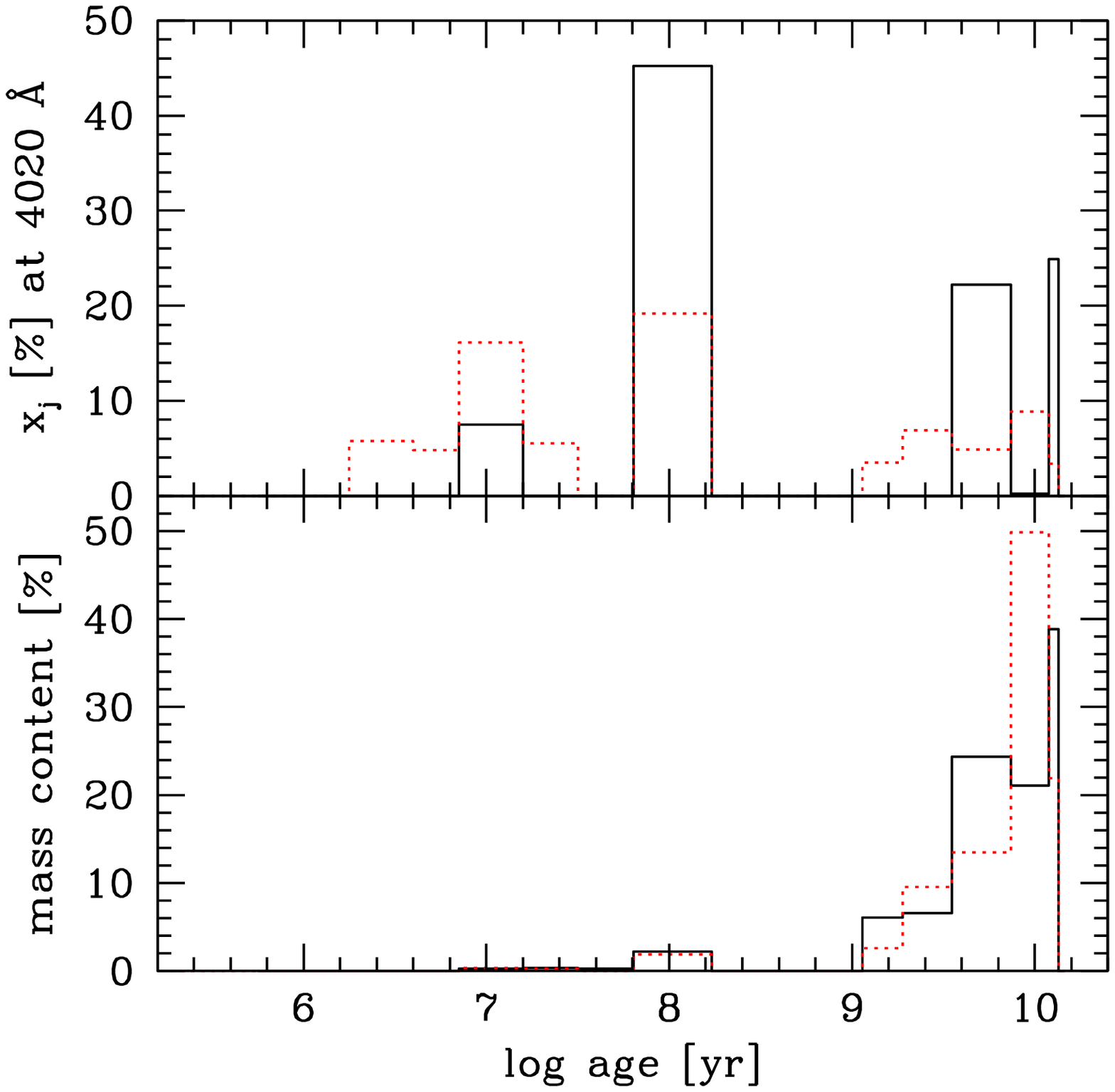}
    \caption{The properties of the stellar populations from a STARLIGHT fit with slope -7/3 (red dotted line; model from the last line of Table~1), and from the complete opt/UV/X-ray fitting (black continuous line) for $a = 0$ (Table~2).}
    \label{fig:populations}
\end{figure}

The properties of the stellar populations from our final fit for $a = 0$ are shown in Fig.~\ref{fig:populations}. We compare it to the results obtained directly from the STARLIGHT code, with a steep power-law model for the disk contribution (see Sect.~\ref{sect:opt/uv_fitting}. Our XSPEC model provided a similar solution, with a somewhat broader distribution of the stellar ages. The new fit still provides qualitatively similar results to the original paper of Bian \& Huang (2010), but not identical. We confirm that two populations of stars are characteristic for the nucleus of RE J1034+396. In our solution, most of the light actually comes from the young stellar population, with ages 10 - 100 million years, and solar metallicity of 0.02. The stars are thus older than implied by the Bian \& Huang fit to the optical data alone. In our case, HST data provides strong constraints on the youngest stellar population. The stars are hidden in the highly obscuring medium, as implied by $A_V = 0.62$. 


\subsection{Spectral features in absorption}

The decomposition of the optical/UV spectrum into the disk and starlight depends on the description of the disk model. Formal solution with a red power law is better in $\chi^2$ term than with a disk-imitating blue power law (see Table~\ref{table:FirstStage}), or a true disk spectra fitted to the broadband continuum (see Table~\ref{tab:optagn_fitting}). Therefore, we take a closer look at individual spectral features to see which of the two models provides better representation of the data. 

A characteristic property of  stellar atmospheres are their absorption features. They form when the radiation from a hot stellar interior passes through the cooler stellar atmospheres. Therefore, the depth of the atomic features (absorption lines and absorption molecular bands, absorption edges) is frequently used to estimate the relative importance of the stellar emission, in comparison to non-stellar emission, e.g. synchrotron radiation. STARLIGHT software, based on the stellar atmosphere models, predicts the existence of such features.
 
Unfortunately, the starlight contamination in an AGN is not easy to determine just from its individual absorption features since some of the usual starlight signatures in the form of absorption lines are filled up with broader emission lines from the hot irradiated gas in the BLR. We first checked directly in the data whether any absorption features are actually seen in our object. We used the features recommended by Cid Fernandes et al. (1998). 

\begin{figure}
    \centering
\includegraphics[width=1.4\hsize]{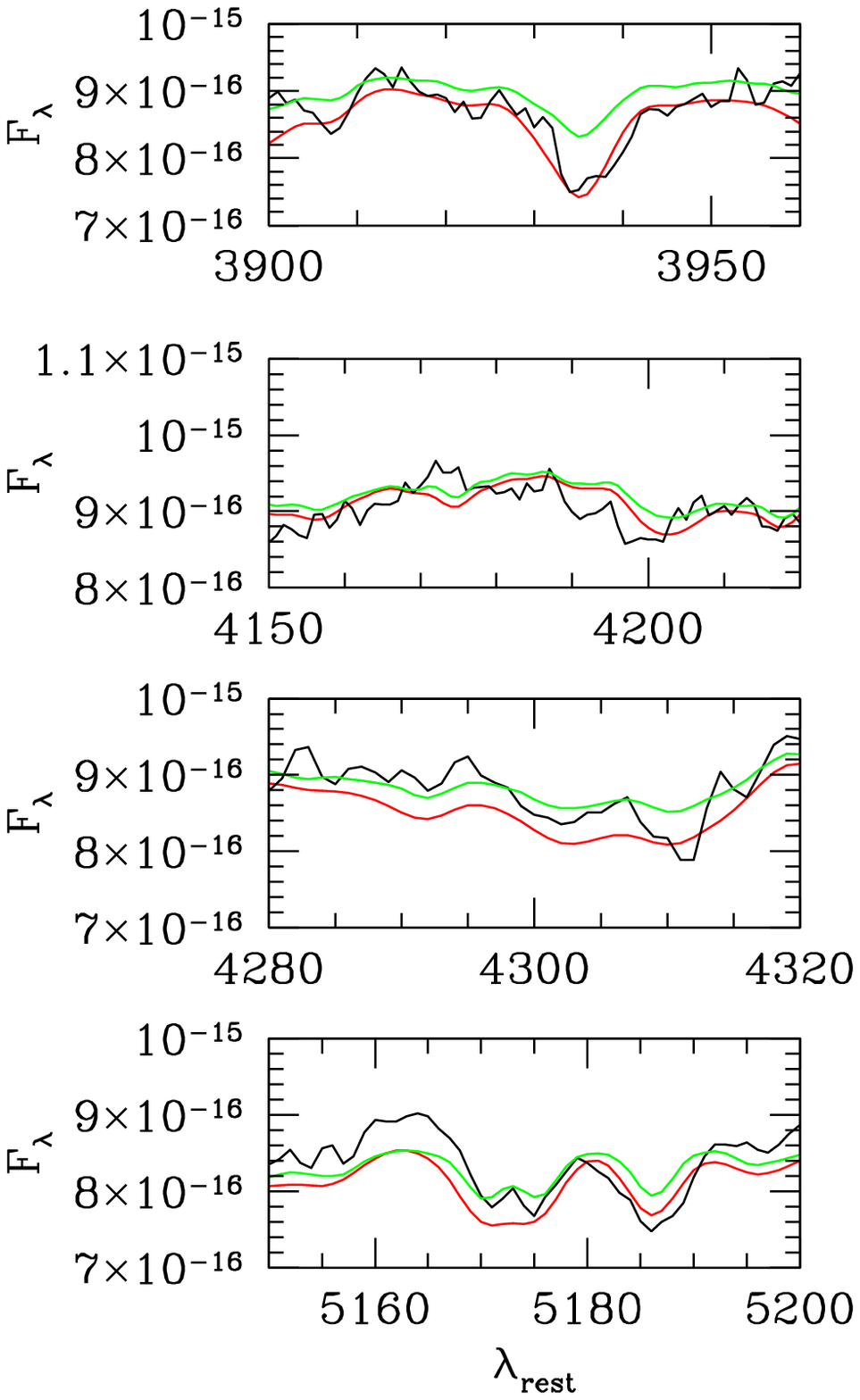}
    \caption{The Ca II K, CN, Gband and Mg I absorption features: the observed spectrum (black line), and best fit power-law solution (green line) and the AGN disk solution $a = 0$ from Table~2 (red line).}
    \label{fig:absorption}
\end{figure}

We searched for these features in the available spectra. The enlarged spectra in the promising wavelength range are shown in Fig.~\ref{fig:absorption}. We did not include the Ca II H + H$\epsilon$ band $\lambda\lambda$3952-3988 since it is strongly contaminated by emission and the Na I region  $\lambda\lambda$5880-5914 since there is no visible spectral feature. We see some traces of absorption in four bands. The Ca II K is the most clear case, additionally located in the bluer part of the spectrum. Here we see that the complex but physically justified model of the accretion disk contribution reproduces  the depth of the feature well while the blue power-law solution (the first line in Table~\ref{table:FirstStage}) gives too shallow an absorption feature meaning that the disk contribution is overestimated in this model.  The other three features are not so deep, and both fits are comparable, although  they formally differ by a factor of 2 in the derived stellar dispersion value. This is because the data, as well as the input models, are marginal for the purpose of the stellar dispersion measurement.

\begin{figure}
    \centering
\includegraphics[width=0.95\hsize]{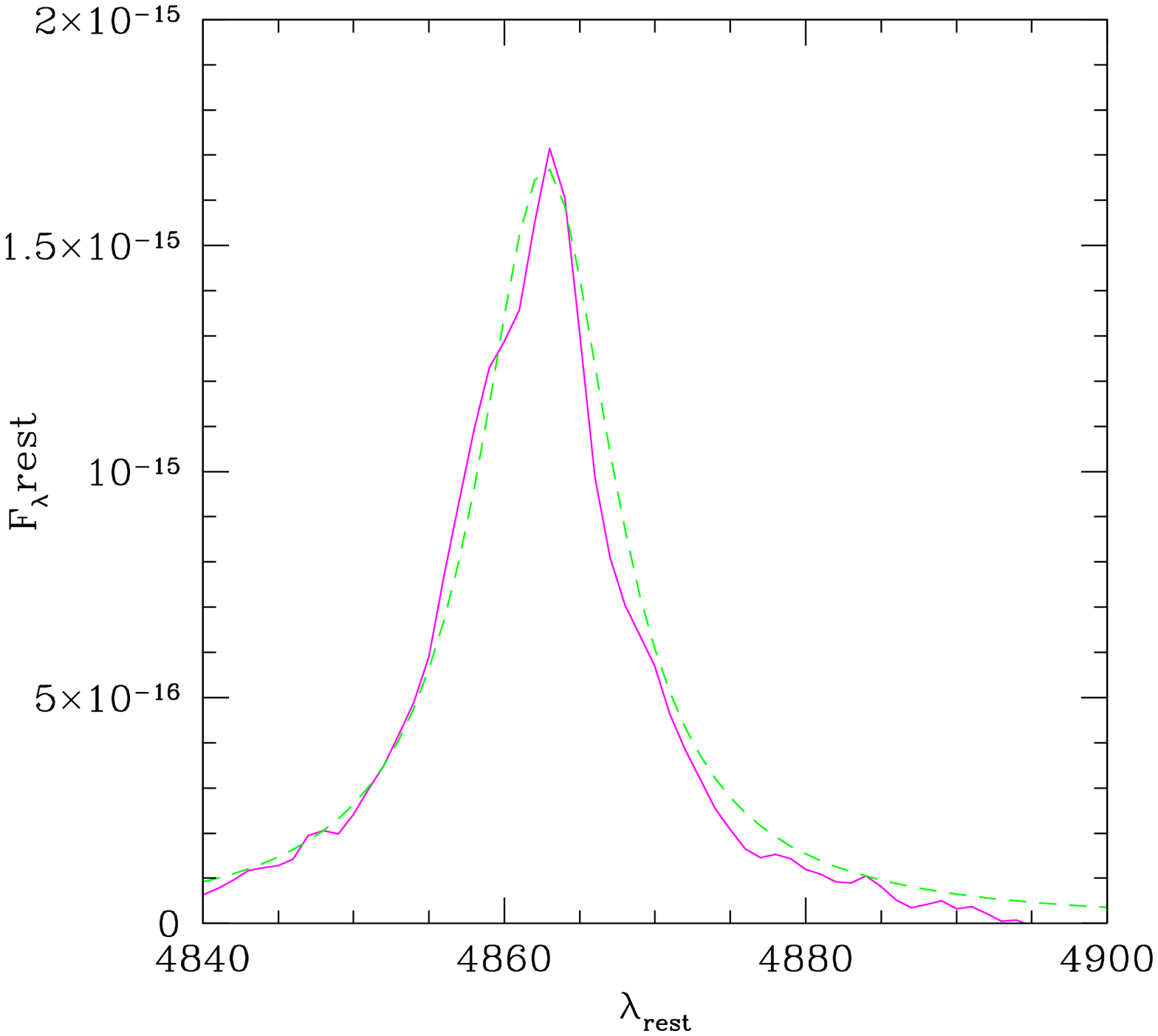}
    \caption{The H$\beta$ line, with starlight and all other contributions subtracted; the observed line (magenta) is quite well-fitted with a single Lorentzian (green dashed line) of FWHM = 5 \AA.}
    \label{fig:Hbeta}
\end{figure}

\subsection{Black hole mass from the stellar dispersion}

The stellar dispersion obtained from the XSPEC code in both cases of $a = 0$ and $a =0.998$ is  $63^{+14}_{-17}$ km s$^{-1}$ (90 \% confidence level). This reduces to true stellar velocity dispersion of $88^{+11}_{-12}$ km s$^{-1}$ when corrected for the spectral resolution of the data and instrumental effects (see Eq. 3 of Bian \& Huang 2010). This value implies the black hole mass $8.5 \times 10^6 M_{\odot}$ if we use the formula (3) of Kormendy \& Ho (2013). The error of the mass measurement is determined by the error of the stellar dispersion, which is not measured accurately for our spectra, so the implied mass is in the range $(4.5 \times 10^6 - 1.4 \times 10^7)M_{\odot}$.  It is higher than the value obtained by Bian \& Huang (2010), $ (1 - 4) \times 10^6 M_{\odot}$, since our measured velocity dispersion in the final models are higher.

\subsection{Black hole mass from H$\beta$ and other emission lines}

The black hole mass can be measured using the emission lines present in the spectrum. Our data set covers two broad lines frequently used for that purpose, H$\beta$, and Mg II, as well as the narrow line [O III]. 
New decomposition of the optical spectrum enable us to revisit the determination of the black hole mass from the H$\beta$ line. The line seems somewhat narrower, full width at the half maximum (FWHM) is 680 km s$^{-1}$, and the line shape is quite well represented just by a single Lorentzian shape (see Fig.~\ref{fig:Hbeta}). The line position coincides well with the vacuum line position (4862.721 \AA), which also supports  our choice of redshift. The Mg II line is somewhat broader, FWHM = 750 km s$^{-1}$, with traces of the doublet structure clearly visible. 

We determine the black hole mass from the formula
\begin{equation}
\log M_{BH} = A + B \log \lambda L_\lambda + 2 \log (FWHM),
\end{equation}
where $M_{BH}$ is in units of $M_{\odot}$, $\lambda L_\lambda$ is in units of $10^{44}$ erg s$^{-1}$, FWHM in 1000 km s$^{-1}$. The coefficients A and B are 6.91 and 0.5 for H$\beta$ (Vestergaard \& Osmer 2009), with the continuum measured at 5100 \AA, and for Mg II they are 6.86 and 0.47, respectively (Vestergaard \& Peterson 2006), and the continuum is measured at 3000 \AA. These values were successfully used in a recent paper by Sun et al. (2015) for a sample of objects. 

The bolometric luminosity of RE J1034+369 is 44.43 or 44.33, depending on the solution, and the monochromatic luminosity at 5100 \AA ~ is 42.70. We note that the bolometric correction for this source is exceptionally large, much larger than the factor of 9 - 10 that is usually adopted. With this value of a monochromatic flux, the formula above  gives $\log M_{BH} = 5.9$ from H$\beta$, and $\log M_{BH} = 6.15$ from Mg II.  

We also try the formula of Collin et al. (2006), their Eq.~7. This formula implies the $f$ factor of 1.83 for H$\beta$, and the corresponding value of the black hole mass $\log M_{BH} = 6.08$. All these values are consistent with each other. 

[O III] line is highly asymmetric. When all other spectral components are subtracted, the red part of the line is consistent with a single Gaussian shape, but the blue part shows strong shoulder. If we use the red part of the line, we obtain the velocity dispersion of 138 km s$^{-1}$. If we correct this value for the instrumental broadening of SDSS data (60 km s$^{-1}$), we obtain 124 km s$^{-1}$. Bian \& Huang (2010) made a proper decomposition of the line and obtained the stellar dispersion from the narrow component of 124 km s$^{-1}$. If this value is treated as an indicator of the stellar dispersion, the black hole mass implied is $2 \times 10^7 M_{\odot}$. We also checked that other forbidden lines, for example the line [O II]$\lambda$3727 \AA,~ is well visible and has similar kinematic width so the use of [O III] seems appropriate. 

\section{Black hole mass from the X-ray variability}

The timescales of the X-ray variability are expected to scale with the size of the region and therefore with the black hole mass. However, the Eddington ratio and other parameters can in principle affect the frequencies and the amplitudes of the variability. Since we do not have a firm model of the dynamics of the X-ray emitting region, the proposed scalings have a phenomenological character. The power spectrum of the X-ray emission in an accreting black hole has a broadband character, most frequently modeled as a broken power law, occasionally with some narrow features (i.e. QPO), classified as Low Frequency QPOs (LF QPO) and High Frequency QPO (HF QPO). HF QPO seen in a number of galactic black holes come in pairs, which correspond to a 2:3 resonance, but  both components of a pair are not always observed.

\subsection{X-ray excess variance}

\subsubsection{Method}

The black hole mass can be conveniently determined from the high frequency tail of the X-ray power spectrum density (e.g. Hayashida et al. 1998, Czerny et al. 2001). The method was inspired by the fact that the PSD in Cyg X-1 extends down to $\sim 25$ Hz with almost the same normalization both in the hard and in the soft states, although a strong evolution with a spectral state takes place at lower frequencies (e.g. Revnivtsev et al. 2000, Gilfanov et al. 2000; Axelsson 2008). Later, it was seen  that X-ray PSD of AGN have a similar shape (e.g. Markowith et al. 2003; Gonzales-Martin \& Voughan 2012) but the timescales are much longer, likely because of the size of the emission region, which is proportional to the black hole mass. So far, there is no strict theory behind the X-ray variability but it's character seems to be universal, and the most plausible mechanism is the picture of propagating perturbations in the accretion disk (Lyubarskii 1997). 

The general dependence of the X-ray PSD on mass and accretion rate allows to use either the frequency break or the normalization of the high frequency tail. In the case of AGN, determination of the frequency break requires long monitoring since the corresponding timescale is of order of a month. The break position depends both on mass and accretion rate, as shown by McHardy et al. (2002).

The tail is easier to determine in AGN, and the tail is also not affected by the Eddington ratio (see Gierlinski et al. 2008 study of galactic sources in various luminosity states), unlike the position of the break (McHardy et al. 2006). The dependence on mass seems universal for most sources, but some NLS1 seemed to be considerable outliers from the predicted normalization. For these few objects the mass determination from the X-ray variability and from the other methods were mismatched up to the factor 20 (Nikolajuk et al. 2009), with the most notable outlier being PG 1211+143. Therefore, for some NLS1, the mass from the standard X-ray variability formula can be too small by a factor of up to 20. Hints of the problem were already seen in Czerny et al. (2001). We thus take this into consideration when analysing RE J1034+396.
Instead of computing the whole power spectrum, it is more convenient to use the excess variance directly calculated from the data (Nikolajuk et al. 2004). 

We used the 28 lightcurves from the XMM satellite in the 2-10 keV band (see Sect.~\ref{sect:data}). The duration of the individual 
exposures varied from 1300 s to 1.07 day, with the median value of about 40 ks. We determined the dimensionless X-ray variance, $\sigma_{exc}$, normalized by the average flux, for each of the curves separately, and we
calculated the individual values of the black hole mass from each lightcurve using the empirical formula
\begin{equation}
M_{BH} = {C T \over \sigma_{exc}^2},
\end{equation}
where $T$ is the duration of the exposure. The constant C is determined from the analysis of the variability of a binary system Cyg X-1 and the determination of its black hole mass.  Since the most recent value for Cyg X-1 black hole mass is $14.8 \pm 1.0 M_{\odot}$ (Orosz et al. 2011a), here we use the value $C =   1.48 M_{\odot}$s$^{-1}$, smaller than in Nikolajuk et al. (2006)  but higher than in Nikolajuk et al. (2009).  The value of the coefficient
C given above is appropriate for Seyfert 1 galaxies, but it may need to be additionally rescaled by a factor of 20 for typical Narrow Line Seyfert 1 galaxies with soft X-ray spectra. Since RE J1034+396 is a typical NLS1 case, we can also use $C = 0.0148  M_{\odot}$s$^{-1}$.

We obtain the final value of the black hole mass by
averaging the $1/M_{BH}$
individual measurements, which roughly corresponds to averaging the variance.  The lowest values of the variance are usually determined
with the largest error, and they are sometimes negative, formally giving very large negative values of the black hole mass. The 
way of
averaging, described above (used also by Nikolajuk et al. 2009), gives less weight to such measurements. The error of the mass is
determined from the dispersion of the individual measurements. 


\subsubsection{Results}

The value of the black hole mass obtained from the X-ray excess variance is $(4.4 \pm 0.6) \times 10^5 M_{\odot}$, if the correction by a factor of 20 is adopted, and the value of $(8.8 \pm 1.3) \times 10^6 M_{\odot}$, if no additional correction is applied. Since RE J1034+396 is a Narrow Line Seyfert 1, the first determination should be more appropriate. On the other hand, we can say, more conservatively, that the true mass from X-ray variability should be between these two values.

\begin{figure}
    \centering
 \includegraphics[width=0.95\hsize]{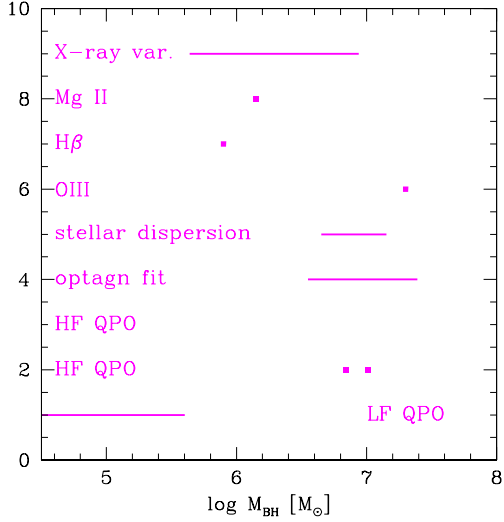}
    \caption{Summary of determinations of the black hole mass in RE J1034+396: from LF or HF higher or lower resonance interpretation of the X-ray QPO, from the broadband fitting of the opt/UV/X-ray spectra with starlight and disk/corona model, from line stellar absorption feature width, from emission lines from NLR ([OIII)] and BLR (H$\beta$ and Mg II), and from the X-ray excess variance method.}
    \label{fig:mas_summary}
\end{figure}

\subsection{QPO}

The phenomenon analogous to the LF QPO have been claimed in several BL Lacs (see King et al. 2013 and the references therein), and the timescale is of the oder of a year. QPO at timescales of hours have been conclusively discovered only in two AGN so far: RE J1034+396 (Gierlinski et al. 2008; see also Alston et al. 2014) and MS 2254.9-3712 (Alston et al. 2015), although there were some other claims in the literature. In galactic sources, QPO are frequently discovered but their duty cycle is also not high, they are seen only in a small fraction of the data sets for a given source. The frequency of the low-frequency component in a given source depends greatly on the source luminosity (see, for example, Homan, Fridriksson \& Remillard 2015). The HF QPO have two peaks (see Remillard \& McClintock 2006 for a review), and their 2:3 ratio implies the resonant character of the phenomenon (Abramowicz \& Klu\' zniak 2001) and the likely scaling with properties of the orbital motion makes them a promising tool for measuring black hole mass and spin (Abramowicz \& Klu\' zniak 2001; Pechacek et al. 2013).

\subsubsection{Method}

If we assume that the QPO observed in RE~J1034+396 is one of the HF QPO, we can use the formula connecting the fundamental frequency with the mass (Remillard \& McClintock 2006)
\begin{equation}
\nu_0 = 931 \left({M \over M_{\odot}}\right)^{-1} {\rm [Hz]}
,\end{equation}
based on the three galactic sources, XTE J1550-564, GRO J1655-40, and GRS 1915+105, which show a pair of HF QPO in 2:3 resonance. Then, if only a single QPO frequency is measured, as here, we have to choose  whether we see the frequency corresponding to 2$\nu_0$ or to 3$\nu_0$.

If we assume that the observed frequencies correspond to the LF QPO we can refer to the scaling with GRS 1915+105, which is the best studied source showing QPO. The study by Yan et al. (2013) shows the LF QPO range in this source between 1 Hz and 8 Hz, depending on the source state. Assuming the value of the black hole mass $14.0 \pm 0.4 M_{\odot}$ in this source (Greiner et al. 2001; Harlaftis \& Greiner 2004), we obtain a minimum and a maximum mass value 
\begin{equation}
M_{min} = { 14.0 \over \nu_0 Hz} M_{\odot}; ~~~~ M_{max} = { 112 \over \nu_0 Hz} M_{\odot},
\end{equation}
for a source displaying a frequency $\nu_0$.

\subsubsection{Results}

Gierlinski et al. (2008) report the detection of the QPO frequency of $2.7 \times 10^{-4}$ Hz, and later observations of RE J1034+396 show a period of $2.6 \times 10^{-4}$ Hz (Alston et al. 2014).

Thus, assuming that the observed QPO corresponds to the higher or lower of the 3:2 resonance 
pair we obtain two mass values: $6.9 \times 10^6 M_{\odot}$ or $1.0 \times 10^7 M_{\odot}$. Adopting the LF QPO interpretation, we obtain the mass range from $5.2 \times 10^4 M_{\odot}$ to $4.2 \times 10^5 M_{\odot}$.

\section{Discussion}

In this paper, we modeled the broad optical/UV/X-ray band of the spectrum of an exceptional Narrow Line Seyfert 1 galaxy RE J1034+396. The optical/UV part of the spectrum is strongly contaminated by starlight emission. This is nicely consistent with no polarization detected in the optical band by Breeveld \& Puchnarewicz (1998). Two separate stellar populations are clearly seen, as obtained before by Bian \& Huang (2010) from the optical data analysis alone. However, the additional use of HST and X-ray data changes the fit quantitatively.

\subsection{Black hole mass}

We used several methods to determine the black hole mass in this source. The summary of the results is shown in Fig.~\ref{fig:mas_summary}. The lines show acceptable ranges, where additional parameters are strongly involved, and the points show individual measurements. 

Some of the methods give a broad range of acceptable values. The broadband fitting of the optical/UV/X-ray spectrum ( {\sc optxagnf} fit) is not unique since the data provides two most stringent constraints: normalization of the disk contribution in optical/UV and the bolometric luminosity, while three essential parameters are needed to model the disk: black hole mass, accretion rate, and spin. We find satisfactory solutions for maximally rotating and non-rotating black hole, which differ in mass by a factor  of 7. The size the corona and the stellar contribution adjusts in such way that the spectra look almost identical. The Eddington ratio and the accretion disk corona size change greatly along the sequence, but the starlight properties barely change. The method of the X-ray excess variance also give a range of black hole mass values if we allow for uncertainty of 20 in the correction factor owing to the AGN type.
Clearly, not all these methods give results which can be accommodated in a single most favorable result. 

Interpretation of the QPO as LF QPO is inconsistent with other measurements. The frequency range of LF QPO in galactic sources is similar in independent measurements (1 - 8 Hz in Yan et al. 2013; 0.5 - 10 Hz in Zhang et al. 2015, in GRS 1915+105). This seems to rule out LF QPO interpretation for the observed QPO phenomenon in this source (Gierlinski et al. 2008, Alston et al. 2014).

The remaining measurements group around two values: $10^6 M_{\odot}$ or $10^7 M_{\odot}$. Higher value is consistent with broadband fitting, stellar dispersion measurement, measurement based on  [O III], high-frequency interpretation of QPO oscillations, and with X-ray variability, if no correction factor is applied for an AGN type.

Lower values are obtained from FWHM of H$\beta$ and Mg II, and from the X-ray variability if the additional scaling factor owing to AGN type is applied. Such small masses imply super-Eddington luminosities since the bolometric luminosity of the source corresponds to the value of the black hole mass  $ \sim 1 \times 10^6 M_{\odot}$.
The measured quantities are certainly affected by some systematic measurement error but it is unlikely that this can be easily reconciled.

The measurement of the stellar dispersion with the SDSS data is not, in our opinion, highly reliable since the dispersion requested by the fitting greatly changes  with the disk description. The measured velocity dispersion, here given as 88 km s$^{-1}$ in final broadband fits, provide the black hole mass value consistent with the value derived from the fit itself. However, the stelar dispersion measured differs considerably for a different description of the optical/UV continuum.  On the other hand, the measurement based on [O III] should be quite precise. This value implies much higher bulge mass than the value expected from the standard relation between the bulge and the central black hole. RE J1034+396 belongs to the NLS1 class, and there is an ongoing discussion whether NLS1 galaxies follow the standard relation or lag behind in their evolution, since they are still at the stage of rapid black hole growth (see e.g. Mathur et al. 2001; Woo et al. 2015). The mass determination also depends on the formula for the relation between the stellar dispersion and black hole mass. In Fig.~\ref{fig:mas_summary}, we used the relation by Kormendy \& Ho (2013). However, if we use the relation of McConnell \& Ma (2013), or of Graham \& Scott (2013), for barred galaxies, as appropriate for RE J1034+396 (Nair \& Abraham 2010), we get the value of the black hole mass $5.5 \times 10^7 M_{\odot}$ and $4.2 \times 10^6 M_{\odot}$, correspondingly, and the uncertainly is large due to the large error of the dispersion measurement. The issue is additionally complicated by the rotational contamination of the stellar features (Woo et al. 2015). Since the black hole mass is small, and the bulge of the galaxy is also small, the measured starlight contains significant light from the disk in this spiral galaxy; this may cause significant systematic error.

With the HF QPO interpretation, both values are then quite precise, and consistent with broad band fitting. The formal fit is somewhat better for lower value, corresponding to twice the fundamental frequency. It is difficult to assess at present the flexibility in the mass values given by Gierlinski et al. (2008). These values come directly from the relation for the fundamental frequency $f_0 = 931 (M/M_{\odot})^{-1}$ Hz (Remillard \& McClintock 2006) coming from three sources XTE J1550-564, GRO J1655–40,
and GRS 1915+105. All of them are microquasars, with black hole spin covering a broad range (XTE J1550-564, $a \sim 0.5$, Steiner et al. 2011; GRS 1915+105, $a \sim 0.95$ (Fragos \& McClintock 2015, You et al. 2015); GRO J1655–40, $ a \sim 0.3$, Motta et al. 2014). The masses of these sources are also determined with some errors, affecting the scaling. The mass of XTE J1550-564 has been revised by a factor of 2 after the work of Remillard \& McClintock (2006) was completed (Orosz et al. 2011b). However, it is rather unlikely that the error is by an order of magnitude. It is interesting to note that, in both cases, the black hole spin must be high, close to the maximally rotating solution. The object does not have a radio detection, and does not show any other imprints of the presence of a jet. On the other hand, there is no firm theory of QPO (see, for example,  Czerny et al. 2010, Czerny \& Das 2011, Hu et al. 2014, Alston et al. 2015 and the references therein). If we try to connect the QPO just to the Keplerian frequency at the innermost stable circular orbit (see e.g. Lei et al. 2015), we  obtain a broad range of possible mass, $ 6.64 \le \log M_{BH} \le 7.69$ values, since the black hole spin in our fitting is unconstrained. However, the discrepancy with the H$\beta$ and Mg II mass measurement remains. Overall, the sources with confirmed HF QPO align well with the Remillard \& McClintock (2006) relation, with RE J1034+396 being a single outlier (Zhou et al. 2015). 

This discussion shows that, despite many efforts, the black hole mass determination in extreme sources remains a challenge. The problem is less severe for typical sources, although Shankar et al. (2016) mention that the mass measurement errors may be partially responsible for the lack of observed correlation between the masses and the AGN spectral shapes, contrary to the theoretical expectations. 

\subsection{Dust component}

 The too shallow slope of the power law found by Bian \& Huang (2010) may be related to the presence of the hot dust emission in the near-IR. The presence of the hot dust in this source has also been postulated by Breevald \& Puchnarewicz (1998) on the basis of detection of highly ionized iron, e.g. [Fe XI]$\lambda$7892.  The studies of AGN in polarized and unpolarized light in the near-IR indeed show that, in the studied quasars, the contribution of the hot dust is important (Kishimoto et al. 2008). However, we did not include this component in our fitting since the model is already quite complex. 

\subsection{Starlight}

In this source the properties of  starlight are exceptionally well measured owing to the very low contribution of the disk to optical/UV band. It is interesting to note the presence of two populations, although the details depend on the disk model, so the young population in Bian \& Huang (2010) is much younger ($\sim$ 10 Myr) than in our solution ($\sim$ 100 Myr). RE J1034+369 is not a starburst galaxy so this kind of a starlight can be treated as a characteristic for all AGN, including quasars. Frequently, the stellar contamination is modeled using only an older stellar population, but this may overestimate the disk contribution at shorter wavelengths. This is particularly important in extremely blue sources, like a quasar PG1426+015 (Watson et al. 2008). 

\section{Conclusions}

We used several black hole mass determination methods for a single source, RE J1034+396. This required broadband spectral modeling in the whole opt/UV/X-ray band, and the detailed analysis of the stellar contribution to the optical/UV spectra. 

\begin{itemize}
\item the methods give contradictory results
\item the methods based on [O III] and stellar dispersion are likely biased
\item broadband fits of the disk/corona model can accommodate a range of masses
\item there is a clear and strong contradiction between the most popular  H$\beta$ method and the HF QPO method; both have formally negligible measurement errors but the results differ by an order of magnitude   
\item the integrated bolometric luminosity implies the black hole mass of $1.0 \times 10^6 M_{\odot}$, if the source radiates at the Eddington luminosity
\item starlight in AGN should be modeled as a two component mixture of older and younger stellar population
\end{itemize}

\begin{acknowledgements}
We are grateful to Ari Laor and  Alister Graham for very helpful comments to the original version of the manuscript, and to the anonymous referee who helped us to improve the paper significantly.
J\'S, KH, BC, MK, BY and A\'{S} 
acknowledge the support by the Foundation for Polish Science through the
Master/Mistrz program 3/2012. The project was partially supported European Union Seventh Framework Program (FP7/2007-2013) under grant
agreement No. 312789; Ministry of Science
and Higher Education grant W30/7.PR/2013 [7.PR] and by the grants of Polish National Science Ceter 2012/04/M/ST9/007870 and 2015/17/B/ST9/03436. The STARLIGHT project is supported by the Brazilian agencies CNP, CAPES and
FAPESP and by the France-Brazil CAPES/Cofecub program. The Fe II theoretical templates described in 
Bruhweiler \& Verner (2008) were downloaded from the
web page http://iacs.cua.edu/personnel/personal-verner-feii.cfm with the permission of the authors. The observational templates of Vestergaard \& Wilkes (2001) were kindly provided by the authors. This research has made use of the NASA/IPAC Extragalactic Database (NED), which is operated by the 
Jet Propulsion Laboratory, California Institute of Technology, under contract with the National Aeronautics and Space Administration.
\end{acknowledgements}


\begin{thebibliography}{99}
\bibitem{}{} Abramowicz, M. A. \& Klu\' zniak, W. 2001, A\&A, 374, L19
\bibitem{}{} Alston, W., Fabian, A., Markeviciute, J., Parker, M., Middleton, M. et al., 2015, MNRAS, 449, 467
\bibitem{}{} Alston, W.N., Markeviciute, J., Kara, E., Fabian, A.C., Middleton, M., 2014, MNRAS, 445, L16
\bibitem{}{} Arnaud, K. A. 1996, Astronomical Data Analysis Software and Systems V, 101, 17 
\bibitem{}{} Aversa, R. et al., 2015, ApJ, 810, 74
\bibitem{}{} Axelsson, M., 2008, AIPC, 1054, 135
\bibitem{}{} Bentz, M.C. et al., 2009a, ApJ, 697, 160
\bibitem{}{} Bentz, M. C., Peterson B. M., Pogge R.W. et al. 2009b, 694, L166
\bibitem{}{} Bentz, M.C. et al., 2010, ApJ, 716, 993
\bibitem{}{} Bentz, M.C. et al., 2014, ApJ, 796, 8
\bibitem{}{} Bian, W.-H., Huang, K., 2010, MNRAS, 401, 507
\bibitem{}{} Bian, W., Zhao, Y., 2004, MNRAS, 352, 823
\bibitem{}{} Boroson, T.A., Green, R.F., 1992, ApJS, 80, 109
\bibitem{}{} Breeveld, A.A., \& Puchnarewicz, E.M., 1998, MNRAS, 295, 568
\bibitem{}{} Bruhweiler, F., \& Verner, E., 2008, ApJ, 675, 83
\bibitem[]{} Bruzual G., Charlot S., 2003, MNRAS, 344, 1000
\bibitem{}{} Cao, X., 2009, MNRAS, 394, 207
\bibitem{}{} Capellupo, D.M. et al., 2015, MNRAS, 446, 3427
\bibitem{}{} Cardelli, J. A., Clayton, G. C., Mathis, J. S., 1989, ApJ, 345, 245     
\bibitem{}{} Casebeer, D. A.. Leighly, K. M., Baron, E., 2006, ApJ, 637, 157
\bibitem[]{} Cid Fernandes, R., Schoenell, W., Gomes, J. M.. Asari, N. V., Schlickmann, M. et al., 2009, RMxAC, 35, 127
\bibitem[]{} Cid Fernandes R., Mateus A., Sodre L., Stasinska G., Gomes J., 2005, MNRAS, 358, 363
\bibitem{}{} Cid Fernandes, R., Storchi-Bergman, T., Schmitt, H.R., 1998, MNRAS, 297, 579
\bibitem{}{} Collin, S., Kawaguchi, T., Peterson, B. M., Vestergaard, M., 2006, A\&A, 456, 75
\bibitem{}{} Crummy J., Fabian A. C., Gallo L., Ross R. R., 2006, MNRAS, 365, 1067
\bibitem{}{} Czerny, B., Hryniewicz, K., Maity, I., Schwarzenberg-Czerny, A., Zycki, P.T., Bilicki, M., 2013, A\&A, 556, A97
\bibitem{}{} Czerny, B., Lachowicz, P., Dovciak, M., Karas, V., Pechacek, T. et al., 2010, A\&A, 524, A26
\bibitem{}{} Czerny, B., Hryniewicz, K., Nikolajuk, M., Sadowski, A., 2011, MNRAS, 415, 2942
\bibitem{}{} Czerny, B., Niko\l ajuk, M., 2010, Memorie della Societa Astronomica Italiana, 81, 281
\bibitem[]{} Czerny, B., Niko\l ajuk, M., R\' o\. za\' nska, A., Dumont, A.-M., Loska, Z., \. Zycki, P.T., 2003, A\&A, 412, 317
\bibitem[]{} Czerny, B., Niko\l ajuk, M., Piasecki, M., Kuraszkiewicz, J., 2001, MNRAS, 325, 865
\bibitem{}{} Davis, S. W., \& Laor, A. 2011, ApJ, 728, 98
\bibitem{}{} Denney, K. D., Peterson, B. M, 2010, ApJ, 721, 715
\bibitem{}{} De Rosa, A., Bianchi, S., Bogdanović, T., Decarli, R., Herrero-Illana, R., et al., 2015, MNRAS, 453, 214 
\bibitem{}{} Dietrich, M., Hamann, F., Shields, J. C., Constantin, A., Vestergaard, M. et al., 2002, ApJ, 581,
912
\bibitem{}{} Done, C., Davis S.W., Jin C., Blaes O.,Ward M., 2012, MNRAS, 420, 1848
\bibitem{}{} Done, C., Gierli\'nski, M., \& Kubota, A., 2007, A\&A Rev., 15, 1
\bibitem{}{} Du, P., Hu, C., Lu, K.-X., Huang, Y.-K., Cheng, C. et al., 2015, ApJ, 806, 22
\bibitem{}{} Du, P., Hu, C., Lu, K.-X., Wang, F., Qiu, J.et al., 2014, ApJ, 782, 45 
\bibitem{}{} Dubois, Y., Volonteri, M.,  Silk, J., 2014, MNRAS, 440, 1590
\bibitem{}{} Edelson, R., et al., 2015, ApJ, 806, 129
\bibitem{}{} Edri, H., Rafter, S.E., Chelouche, D., Kaspi, S., Behar, E., 2012, ApJ, 756, 73
\bibitem{}{} Fabian, A.C. et al., 2015, MNRAS, 451, 4375
\bibitem{}{} Fragos, T., McClintock, J.E., 2015, ApJ, 800, 17
\bibitem{}{} Gaskell, C. M. 2009, arXiv:0908.0328
\bibitem{}{} Gilfanov M., Churazov E., Revnivtsev M., 2000, MNRAS, 316, 923
\bibitem{}{} Gierli\' nski, M., Middleton, M., Ward, M., Done, C., 2008, Natur, 455, 369
\bibitem{}{} Gierli\' nski, M., Niko\l ajuk, M., Czerny, B., 2008, MNRAS, 383, 741
\bibitem{}{} Gonzales-Martin, O., Vaughan, S., 2012, A\&A, 544, A80
\bibitem{}{} Grandi, S. A., 1982, ApJ, 255, 25
\bibitem{}{} Graham, A. W., Scott, N., 2013, ApJ, 764, 151
\bibitem{}{} Greene, J.E., Ho, L.C., 2005, ApJ, 630, 122
\bibitem{}{} Greiner, J., Cuby, J., McCaughrean, M. 2001, Nature, 414, 522
\bibitem{}{} Grier, C. et al., 2012, ApJ, 755, 60
\bibitem{}{} Grier, C. et al., 2013, ApJ, 773, 90
\bibitem{}{} Harlaftis, E. T.,  Greiner, J., 2004, A\&A, 414, L13
\bibitem{}{} Hayashida, K., Miyamoto, S., Kitamoto, S., Negoro, H., Inoue, H., 1998, ApJ, 504, L71
\bibitem{}{} Homan, J., Fridriksson, J. K., Remillard, R. A., 2015, ApJ, 812, 80
\bibitem{}{} Hryniewicz, K. et al., 2014, A\&A, 562, A34
\bibitem{}{} Hu, C. et al., 2014, ApJ, 788, 31
\bibitem{}{} Hu, C. et al., 2015, ApJ, 804, 138
\bibitem{}{} Karachentsev, I., Lebedev, V., \& Shcherbanovskij, A., 1985, Bull. Inform. CDS, 29, 87 
\bibitem{}{} Kaspi, S. et al., 2000, ApJ, 533, 631
\bibitem{}{} Kishimoto, M., Antonucci, R., Blaes, O., Lawrence, A., Boisson, C. et al., 2008, Natur, 454, 492
\bibitem{}{} King, A.L.  et al., 2015, MNRAS 453, 1701
\bibitem{}{} King, O. G., Hovatta, T., Max-Moerbeck, W., Meier, D. L., Pearson, T. J. et al., 2013, MNRAS, 436, L114
\bibitem{}{} Kollmeier, J.A. et al., 2006, ApJ, 648, 128
\bibitem{}{} Kormendy, J., \& Ho, L. C. 2013, ARA\&A, 51, 511
\bibitem{}{} Kuraszkiewicz, J. et al., 2009, ApJ, 692, 1180
\bibitem{}{} Lei, W.-H., Yuan, Q., Zhang, B. \& Wang, D., 2015, ApJ, 816, 20
\bibitem{}{} Loska, Z., Czerny, B., Szczerba, R., 2004, MNRAS, 355, 1080
\bibitem{}{} Lyubarskii, Yu.E., 1997, MNRAS, 292, 679
\bibitem{}{} Maitra, D., Miller, J. M., 2010, ApJ, 718, 551
\bibitem{}{} Markowith, A., Edelson, R., Vaughan, S., Uttley, P. George, I.M. et al.,, 2003, ApJ, 593, 96
\bibitem{}{} Marziani, P., Sulentic, J. W., Zwitter, T., Dultzin-Hacyan, D., \& Calvani, M. 2001, ApJ, 558, 553
\bibitem{}{} Mason, K.O., Puchnarewicz, E.M., Jones, L.R., 1996, MNRAS, 283, L26
\bibitem{}{} Mathur, S., Kuraszkiwicz, J., Czerny, B, 2001, New Astronomy, 6, 321
\bibitem{}{} McConnell, N.J., Ma, C.-P., 2013, ApJ, 764, 184
\bibitem{}{} McHardy, I. M., Koerding, E., Knigge, C., Uttley, P., Fender, R. P., 2006, Natur, 444, 730
\bibitem[]{} Middleton, M.,  Uttley, P., \& Done, C., 2011, MNRAS, 417, 250
\bibitem{}{} Middleton, M., Done, C., Ward, M., Gierli\' ski, M., Schurch, N., 2009, MNRAS, 394, 250
\bibitem{}{} Modzelewska, J. et al., 2014, A\&A, 570, A53
\bibitem{}{} Motta,  S.  E.,  Belloni,  T.  M.,  Stella,  L.,  Munoz-Darias,  T.,  Fender,  R.  2014, MNRAS, 437, 3, 2554
\bibitem{}{} Nair, P. B., Abraham, R. G., 2010, ApJS, 186, 427
\bibitem{}{} Niko\l ajuk, M., Papadakis, I. E., Czerny, B., 2004, MNRAS, 350, L26
\bibitem{}{} Niko\l ajuk, M., Czerny, B., Zi\' \l kowski, J., Gierli\' nski, M., 2006, MNRAS, 370, 1534
\bibitem{}{} Niko\l ajuk, M., Czerny, B., Gurynowicz, P., 2009, MNRAS, 394, 2141
\bibitem{}{} Novikov, I. D., \& Thorne, K. S. 1973, Black Holes (Les Astres Occlus), 343 
\bibitem{}{} Onken, C. A., Ferrarese, L., Merritt, D., et al. 2004, ApJ, 615, 645
\bibitem{}{} Orosz, J.A. et al., 2011a, ApJ, 742, 84
\bibitem{}{} Orosz J. A., Steiner J. F., McClintock J. E., Torres M. A. P., Remillard R. A., Bailyn C. D., Miller J. M., 2011, ApJ, 730, 75
\bibitem{}{} Pechacek T., Goosmann, R. W., Karas, V., Czerny, B., Doovciak, M. 2013, A\&A, 556, A77
\bibitem{}{} Peng, C.Y., Impey, C.D., Ho, L.C., Barton, E.J. \& Rix, H.-W., 2006, ApJ, 640, 114
\bibitem{}{} Peterson, B.M., 1993, PASP, 105, 247
\bibitem{}{} Peterson, B.M. et al., 2004, ApJ, 613, 682
\bibitem{}{} Peterson, B.M. et al., 2013, ApJ, 779, 109
\bibitem{}{} Peterson, B.M. et al., 2014, ApJ, 795, 149
\bibitem{}{} Pozo Nuñez, F., Ramolla, M., Westhues, C., Haas, M., Chini, R. et al., 2015,  A\&A, 576, A73
\bibitem{}{} Puchnarewicz, E.M., Mason, K.O., Siemiginowska, A., Pounds, K.A., 1995, MNRAS, 276, 20 
\bibitem{}{} Puchnarewicz, E.M., Mason, K.O., Siemiginowska, 1998, MNNRAS, 293, L52
\bibitem{}{} Puchnarewicz, E. M., Mason, K. O., Siemiginowska, A., Fruscione, A., Comastri, A., Fiore, F, \& Cagnoni, I., 2001, ApJ, 550, 644
\bibitem{}{} Rafter, S.E., Kaspi, S., Chelouche, D., Sabach, E., Karl, D. et al., 2013, ApJ, 773, 24 
\bibitem{}{} Remillard, R.A., \& McClintock, J.E., 2006, ARA\&A, 44, 49
\bibitem{}{} Revnivtsev M., Gilfanov M., Churazov E., 2000, A\&A, 363, 1013
\bibitem{}{} R\' o\. za\' nska, A., Malzac, J., Belmont, R., Czerny, B., Petrucci, P.-O., 2015, A\&A, 580, A77
\bibitem{}{} Schlafly, E. F., \& Finkbeiner, D. P. 2011, ApJ, 737, 103
\bibitem[]{} Shakura, N.I., \&  Sunyaev, R.A., 1973, A\&A, 24, 337 
\bibitem{}{} Shankar, F. 2009, NewAR, 53, 57
\bibitem{}{} Shankar, F. et al., 2016, ApJL, 818, L1 
\bibitem{}{} Shen, Y. et al., 2011, ApJS, 194, 45
\bibitem{}{} Shen, Y., Ho, L. C., 2014, Nature, 513, 210
\bibitem{}{} Shen, Y., Brandt, W. N., Dawson, K. S., Hall, P. B., McGreer, I. D. et al., 2015, ApJS, 216, 4 
\bibitem{}{} Shen, Y. et al., 2015, ApJ, 805, 96
\bibitem{}{} Soria, R. Puchnarewicz, E.M., 2002, MNRAS, 329, 456
\bibitem{}{} Steiner, J.F. et al., 2011, MNRAS, 416, 941
\bibitem{}{} Storey, P. J., Hummer, D. G., 1995, MNRAS, 272, 41
\bibitem{}{} Sun, M., Trump, J.R., Brandt, W.N., Luo, B., Alexander, D.M. et al., 2015, 802, 14
\bibitem[]{} Sulentic, J.W., Zwitter, T., Marziani, P., \& Dultzin-Hacyan, D. 2000, ApJ, 536, L5
\bibitem{}{} Titarchuk, L., 1994, ApJ, 434, 313
\bibitem{}{} Valenti, S., Sand, D. J., Barth, A. J., Horne, K., Treu, T. et al., 2015, ApJ, 813, L36
\bibitem{}{} Vestergaard, M.,  \& Osmer, P. S. 2009, ApJ, 699, 800 
\bibitem{}{} Vestergaard, M., Peterson, B.M., 2006, ApJ, 641, 689
\bibitem{}{} Vestergaard, M. \& Wilkes B. J., 2001, ApJS, 134, 1
\bibitem{}{} Watson, L.C., Martini, P., Dasyra, K. M., Bentz, M. C., Ferrarese, L., Peterson, B. M. et al., 2008, ApJ, 682, L21
\bibitem{}{} Wandel, A., Peterson, B.M., Malkan, M.A., 1999, ApJ, 526, 579
\bibitem{}{} Wang, J.-M., Netzer, H., 2003, A\&A, 398, 927
\bibitem{}{} Woo, J.-H., Treu, T., Malkan, M., Blandford, R.D., 2008, ApJ, 681, 925
\bibitem{}{} Woo, J.-H., Yoon, Y., Park, S., Park, D., Kim, S.C., 2015, ApJ, 801, 38
\bibitem{}{} Yan, S.-P. et al., 2013, MNRAS, 434, 59
\bibitem{}{} You, B., Czerny, B., Sobolewska, M. et al.: 2015, arXiv:1506.03959 
\bibitem{}{} You, B., Cao, X., \& Yuan, Y.-F., 2012, ApJ, 761, 109
\bibitem{}{} Zdziarski, A.A.,  Johnson, W.N. \& Magdziarz, P., 1996, MNRAS, 283, 193
\bibitem{}{} Zhou, X.-L., Yuan, W., Pan, H.-W., Liu, Z., 2015, ApJ, 798, L5
\bibitem{}{} Zoghbi, A., Fabian, A.C., 2011, MNRAS, 418, 2642
\bibitem{}{} Zwicky, F., \& Herzog, E., 1966, Catalogue of Galaxies and of Clusters of Galaxies, volume III, Pasadena: California Institute of Technology
\bibitem{}{} Zhang, L., Chen, L., Qu, J.-L., Bu, Q.-C., 2015, AJ, 149, 82
\end{thebibliography}
\end{document}